\documentclass[longauth, oneside]{aa} 

\usepackage{txfonts}
\usepackage{graphicx}   
\usepackage{amsmath}    
\usepackage{amssymb}    
\usepackage{makecell}
\usepackage{subfigure}

\usepackage{natbib} 
\usepackage{color}
\bibpunct{(}{)}{;}{a}{}{,}
\usepackage[flushleft]{threeparttable}
\usepackage{hyperref}

\definecolor{orange}{rgb}{0.99, 0.55, 0.01}

\usepackage[normalem]{ulem}

\begin{document} 

   \title{Probing the atmosphere of WASP-69~b with \\ low- and high-resolution transmission spectroscopy}

   \author{S.~Khalafinejad\inst{\ref{1lsw},\ref{2mpia}}
   \and K.~Molaverdikhani\inst{1,2} 
   \and J.~Blecic\inst{3,4} 
   \and M.~Mallonn\inst{5} 
   \and L.~Nortmann\inst{6}
   \and J.\,A.~Caballero\inst{7}
   \and H.~Rahmati\inst{8}
   \and A.~Kaminski\inst{1} 
   \and S.~Sadegi\inst{1,2}
   \and E.~Nagel\inst{9,10}
   \and L.~Carone\inst{2}
   \and P.\,J.~Amado\inst{11}
   \and M.~Azzaro\inst{12} 
   \and F.\,F.~Bauer\inst{11}
   \and N.~Casasayas-Barris\inst{13}
   \and S.~Czesla\inst{9}
   \and C.~von~Essen\inst{14,15}
   \and L.~Fossati\inst{16}
   \and M.~G\"udel\inst{17}
   \and Th.~Henning\inst{2}
   \and M.~L\'opez-Puertas\inst{11}
   \and M.~Lendl\inst{16,18} 
   \and T.~L\"uftinger\inst{17,19}  
   \and D.~Montes\inst{20}
   \and M.~Oshagh\inst{21,22}
   \and E.~Pall\'e\inst{21,22}
   \and A.~Quirrenbach\inst{1}
   \and S.~Reffert\inst{1}   
   \and A.~Reiners\inst{6}
   \and I.~Ribas\inst{23,24}
   \and S.~Stock\inst{1}
   \and F.~Yan\inst{6} 
   \and M.\,R.~Zapatero~Osorio \inst{25}
   \and M.~Zechmeister\inst{6}
          }
    \institute{Landessternwarte, Zentrum f\"{u}r Astronomie der Universit\"{a}t Heidelberg, K\"{o}nigstuhl 12, 69117 Heidelberg, Germany\\ \email{skhalafinejad@lsw.uni-heidelberg.de} \label{1lsw}
    \and Max-Planck-Institut f\"{u}r Astronomie, K\"{o}nigstuhl 17, 69117 Heidelberg, Germany \label{2mpia}
    \and Department of Physics, New York University Abu Dhabi, PO Box 129188, Abu Dhabi, UAE  \label{3newyork}
        \and Center for Astro, Particle and Planetary Physics (CAP3), New York University Abu Dhabi, PO Box 129188, Abu Dhabi, UAE  \label{4CAP3}
    \and Leibniz-Institut f\"ur Astrophysik Potsdam,
    An der Sternwarte 16, 14482 Potsdam, Germany \label{5AIP}
    \and Institut f\"ur Astrophysik, Georg-August-Universit\"at G\"ottingen, 
    Friedrich-Hund-Platz 1, 37077 G\"ottingen, Germany \label{6Goett}
    \and Centro de Astrobiolog\'ia (CSIC-INTA), ESAC, Camino bajo del castillo s/n, 28692 Villanueva de la Ca\~nada, Madrid, Spain \label{7CAB-Villafranca}
    \and Department of Physics, Bu-Ali Sina University, Hamedan 65178, Iran \label{8Bu-Ali}
    \and Th\"uringer Landessternwarte Tautenburg, Sternwarte 5, 07778 Tautenburg, Germany \label{9Thueringer}
    \and Hamburger Sternwarte, Universit\"at Hamburg, Gojenbergsweg 112, 21029 Hamburg, Germany \label{10Hamburger} 
    \and Instituto de Astrof\'isica de Andaluc\'ia (IAA-CSIC), Glorieta de la Astronom\'ia s/n, 18008 Granada, Spain \label{11IAA}
    \and Centro Astron\'omico Hispano-Alem\'an, Observatorio de Calar Alto, 04550 G\'ergal, Almer\'ia, Spain \label{12CAHA}
    \and Sterrewacht Leiden, Universiteit Leiden, Postbus 9513, 2300 RA, Leiden, The Netherlands \label{13Leiden}
    \and Stellar Astrophysics Centre, Aarhus Universitet, Ny Munkegade 120, 8000 Aarhus C, Denmark \label{14Aarhus}
    \and Astronomical Observatory, Institute of Theoretical Physics and Astronomy, Vilniaus universitetas, Sauletekio av. 3, 10257, Vilnius, Lithuania \label{15Vilnus}
    \and Institut f\"ur Weltraumforschung, \"Osterreichische Akademie der Wissenschaften,
    Schmiedlstrasse 6, A-8042 Graz, Austria \label{16Graz}
    \and Institut f\"ur Astronomie, Universit\"at Wien, T\"urkenschanzstrasse 17, 1180 Wien, Austria \label{17Vienna}
    \and Observatoire de Gen\`eve, Universit\'e de Gen\`eve, Chemin des maillettes 51, 1290 Sauverny, Switzerland \label{18Geneva}
    \and European Space Research and Technology Centre, 
    European Space Agency, Keplerlaan 1, 2201, AZ Noordwijk, The Netherlands \label{19ESTEC}
    \and Facultad de Ciencias F\'isicas, Departamento de F\'isica de la Tierra y Astrof \'isica \& IPARCOS-UCM (Instituto de F\'isica de Part\'iculas y del Cosmos de laUCM), Universidad Complutense de Madrid, 28040 Madrid, Spain \label{20laUCM}
    \and Instituto de Astrof\'isica de Canarias, c/ V\'ia L\'actea s/n, 38205 La Laguna, Tenerife, Spain \label{21IAC}
    \and Universidad de La Laguna, Departamento de Astrof\'isica, E-38206 La Laguna, Tenerife, Spain \label{22ULL}
    \and Institut de Ci\'encies de l’Espai (CSIC-IEEC), Campus UAB, c/de Can Magrans s/n, 08193 Bellaterra, Barcelona, Spain \label{23ICE}
    \and Institut d’Estudis Espacials de Catalunya (IEEC), 08034 Barcelona, Spain \label{24IEEC}
    \and Centro de Astrobiolog\'ia (CSIC-INTA), Carretera de Ajalvir km~4, Torrejón de Ardoz, 28850 Madrid, Spain \label{25CAB-Torrejon}
    }

   \date{Received dd April 2021 / Accepted dd Month 2021}

  \abstract{Consideration of both low- and high-resolution transmission spectroscopy is key for obtaining a comprehensive picture of exoplanet atmospheres. In studies of transmission spectra, the continuum information is well established with low-resolution spectra, while the shapes of individual lines are best constrained with high-resolution observations.
  In this work, we aim to merge high- with low-resolution transmission spectroscopy to place tighter constraints on physical parameters of the atmospheres.
  We present the analysis of three primary transits of WASP-69~b in the visible (VIS) channel of the CARMENES instrument 
  and perform a combined low- and high-resolution analysis using additional data from HARPS-N, OSIRIS/GTC, and WFC3/{\em HST}  already available in the literature.
  We investigate the Na~{\sc i} D$_1$ and D$_2$ doublet, H${\alpha}$, the Ca~{\sc ii} infra-red triplet (IRT), and K~{\sc i} $\lambda$7699\,{\AA} lines, and we monitor the stellar photometric variability by performing long-term photometric observations with the STELLA telescope.
  During the first CARMENES observing night, we detected the planet Na~{\sc i} D$_{2}$ and D$_{1}$ lines at $\sim 7\sigma$ and $\sim 3\sigma$ significance levels, respectively. 
  We measured a D$_{2}$/D$_{1}$ intensity ratio of 2.5$\pm$0.7, which is in agreement with previous HARPS-N observations.
  Our modelling of WFC3 and OSIRIS data suggests strong Rayleigh scattering, solar to super-solar water abundance, and a highly muted Na feature in the atmosphere of this planet, in agreement with previous investigations of this target.
  We use the continuum information retrieved from the low-resolution spectroscopy as a prior to break the degeneracy between the Na abundance, reference pressure, and thermosphere temperature for the high-resolution spectroscopic analysis. 
  We fit the Na~{\sc i} D$_{1}$ and D$_{2}$ lines individually and find that the posterior distributions of the model parameters 
  agree with each other within 1$\sigma$. 
  Our results suggest that local thermodynamic equilibrium processes can explain the observed D$_{2}$/D$_{1}$ ratio because the presence of haze opacity mutes the absorption features.}

\keywords{
methods: observational --
techniques: spectroscopic --
planets and satellites: atmospheres, composition, individual: WASP-69~b --
stars: activity}
   
\maketitle
%
\section{Introduction}

\label{sec:intro}

\begin{table*}
\centering
\label{tbl:orbital parameters}
\caption{Orbital and physical parameters of the WASP-69 star-planet system$^{a}$.}
\begin{tabular}{ l l c }
\hline\hline
\noalign{\smallskip}
Parameter & Symbol (unit) & Value  \\
\hline
\noalign{\smallskip}
        Stellar effective temperature  & $T_{\rm eff}$ (K)    & 4715 $\pm$ 50           \\
        Stellar surface gravity        & $\log{g_{*}}$ (cgs)          & 4.535 $\pm$ 0.023  \\
        Stellar metallicity            & $[\text{Fe/H}]$      &  0.144 $\pm$ 0.077       \\
        Stellar radius                 & $R_{*}$ ($R_{\odot}$)  & 0.813 $\pm$ 0.028    \\ 
        Projected stellar rotation velocity       & $v \sin{i}$ (km\,s$^{-1}$)      & 2.2 $\pm$ 0.4  \\    
        Systemic radial velocity       & $\gamma$ (km\,s$^{-1}$)      & --9.6 $\pm$ 0.2  \\    
\noalign{\smallskip}
\hline
\noalign{\smallskip}
        Mid-transit time               & $T_{0}$ (HJD$_{\text{UTC}}$) & 2455748.83344 $\pm$ 0.00018 \\
        Orbital period                 & $P$ (d)     & 3.8681382 $\pm$ 0.0000017      \\
        Transit duration               & $T_{14}$ (d)  &    0.0929 $\pm$ 0.0012 \\
        Ingress/egress duration        & $T_{12} = T_{34}$ (d)   &  0.0192 $\pm$ 0.0014  \\
        Orbital inclination            & $i$ (deg)      & 86.71 $\pm$  0.20     \\
        Semi-major axis                & $a$ (au)          & 0.04525 $\pm$ 0.00053      \\
        Planet mass                    & $M_{\rm p}$ (M$_{\rm Jup}$)  & 0.260 $\pm$0.017  \\
        Planet radius                  & $R_{\rm p}$ (R$_{\rm Jup}$)  & 1.057 $\pm$ 0.047   \\
        Planet surface gravity         & log $g_{\rm p}$ (cgs)         & 2.726 $\pm$ 0.046 \\

        Planet to star area ratio    & ($R_{\rm p}$/$R_{*}$)$^{2}$     & 0.01786  $\pm$ 0.00042 \\ 
        Planetary equilibrium temperature & $T_{\rm eq}$  ($A_{\rm Bond}$ = 0.0, 0.3, 0.6) (K) & (964, 882, 767)  \\
\noalign{\smallskip}
\hline
\end{tabular}
\tablefoot{
\tablefoottext
{a}{To constrain possible planetary equilibrium temperatures, we performed analytical calculations \citep{Seager2005} assuming efficient heat circulation and different Bond albedos,~$A_{\rm Bond}$.
All other values are adopted from \citet{Anderson2014}. }
}
\end{table*}

The transmission spectroscopy method is applicable when an exoplanet crosses the disk of its host star as seen from Earth. 
Within this geometry, a part of the stellar light passes through the exoplanet atmosphere, and the atmosphere imprints its signature on  the spectrum. 
Through low- and high-resolution spectroscopy, one can access embedded spectral information from different pressure levels in the atmosphere. 
Low-resolution transmission spectra probe deeper levels of the atmosphere; they usually contain information regarding the atmospheric temperature, abundances, the presence or absence of haze and clouds, and the general shape of the atomic and molecular continuum features \citep[e.g.,][]{Sing2008,Vidal-Madjar2011,Fischer2016}. 
On the other hand, high opacities at lower atmospheric pressures can imprint their spectral signatures on high-resolution spectra, resulting in unambiguous detection of atmospheric atomic and molecular features \citep[e.g.,][]{Hoeijmakers2019, Yan2019, Lopez2019, Nortmann2018, Khalafinejad2017}. 
Therefore, a combined or even parallel analysis of both low- and high-resolution spectra provides the complementary information \citep[e.g.,][]{Brogi2017,Pino2018, Carone2021} required to fully unveil exoplanet atmospheres.

\object{WASP-69~b} is a hot, inflated Saturn-mass exoplanet orbiting the active K-type star \object{BD--05~5432} with an orbital period of $\sim$3.87\,d \citep{Anderson2014}. 
The most relevant parameters of this star--planet system are listed in Table~\ref{tbl:orbital parameters}. 
Following the discovery of the planet, there have been several attempts to characterize the atmosphere of WASP-69~b using both high- and low-resolution  instruments. 
\citet{Casasayas-Barris2017} investigated the sodium absorption using high-resolution spectra collected with the High Accuracy Radial velocity Planet Searcher-North (HARPS-N) at the 3.56\,m Telescopio Nazionale Galileo. 
Through comparison between the in-transit and out-of-transit spectra, these latter authors detected an excess absorption only at the Na D$_2$ line in the passband of 1.5\,{\AA}, at the level of $5\sigma$. 
Later, \citet{Deibert2019} searched for sodium, potassium, and water in the atmosphere of WASP-69~b by cross-correlating model spectra with the observed transit spectra taken with the Gemini Remote Access to CFHT ESPaDOnS Spectrograph (GRACES). 
Their analysis did not show any significant signal ($>3\sigma$) of these species. 
Additionally, using the high-resolution transit spectra taken with the CARMENES instrument, \citet{Nortmann2018} detected excess absorption of the helium triplet in the near-infrared regime (He~{\sc i} $\lambda$10830\,\AA) with a signal-to-noise ratio (S/N) of 18.
These authors also observed a post-transit absorption together with a line blueshift of several kilometers per second, which they interpreted as escape from the upper atmospheric layers. 
Recently, \citet{Murgas2020-WASP69b} obtained optical broad-band spectroscopy of WASP-69~b by conducting transit observations with  
the Optical System for Imaging and low-Intermediate-Resolution Integrated Spectroscopy (OSIRIS) at the 10.4\,m Gran Telescopio Canarias (GTC). 
Their transmission spectrum did not show any signs of excess absorption at the sodium doublet wavelengths. However, the slope of the spectrum obtained by these latter authors was consistent with a hazy atmosphere that causes atmospheric scattering.
In addition to these ground-based studies, \citet{Tsiaras2018} used the Wide Field Camera 3 (WFC3) at the {Hubble Space Telescope} ({\em HST}) in the near-infrared and homogeneously studied a sample of giant exoplanets, including WASP-69~b. 
Their low-resolution transmission spectra revealed water vapor features.

In the present paper, we use high-resolution transit spectra of CARMENES (the same observations as \citealt{Nortmann2018} plus a new observation from August 2020) and mainly search for exoplanet atmospheric features in the visible region of the spectrum (e.g., sodium, potassium, and excited atomic hydrogen). 
To increase the S/N of the transmission spectrum, we combine our final results from CARMENES with those of HARPS-N \citep{Casasayas-Barris2017}. We also compare the three individual nights of CARMENES observations and evaluate the level of stellar activity.
To obtain a more complete picture of the atmosphere, we perform atmospheric retrieval modeling on both low- and high-resolution transmission spectra.
For the low-resolution part, we make use of the transmission spectra of WFC3/{\em HST} \citep{Tsiaras2018} and OSIRIS/GTC \citep{Murgas2020-WASP69b} already available in the literature, and look for water features in infrared (IR).

The structure of the paper is as follows. 
In Sect.~\ref{sec:obs}, we present the high-resolution transit observations with CARMENES and describe our initial data-reduction procedure. 
We also describe our long-term photometric observations and analysis with the STELLA telescope. 
In Sect.~\ref{sec:TS}, we analyze the high-resolution transmission spectra around prominent atomic features in the optical and near-infrared spectral regions and we discuss the stellar chromospheric activity.
At the end of this section, we discuss our observational results and a measure of the statistical robustness of our signal. 
In Sect.~\ref{sec:modeling}, we perform atmospheric retrievals on both low- and high-resolution data and discuss our findings. 
Finally, in Sect.~\ref{sec:conclusions} we summarize our findings and provide our conclusions.

\section{Observations and data reduction}
\label{sec:obs}

\subsection{Transit spectroscopy with CARMENES}
\label{sec:obs_carmenes}

\begin{figure}
    \centering
    \includegraphics[width =\columnwidth]{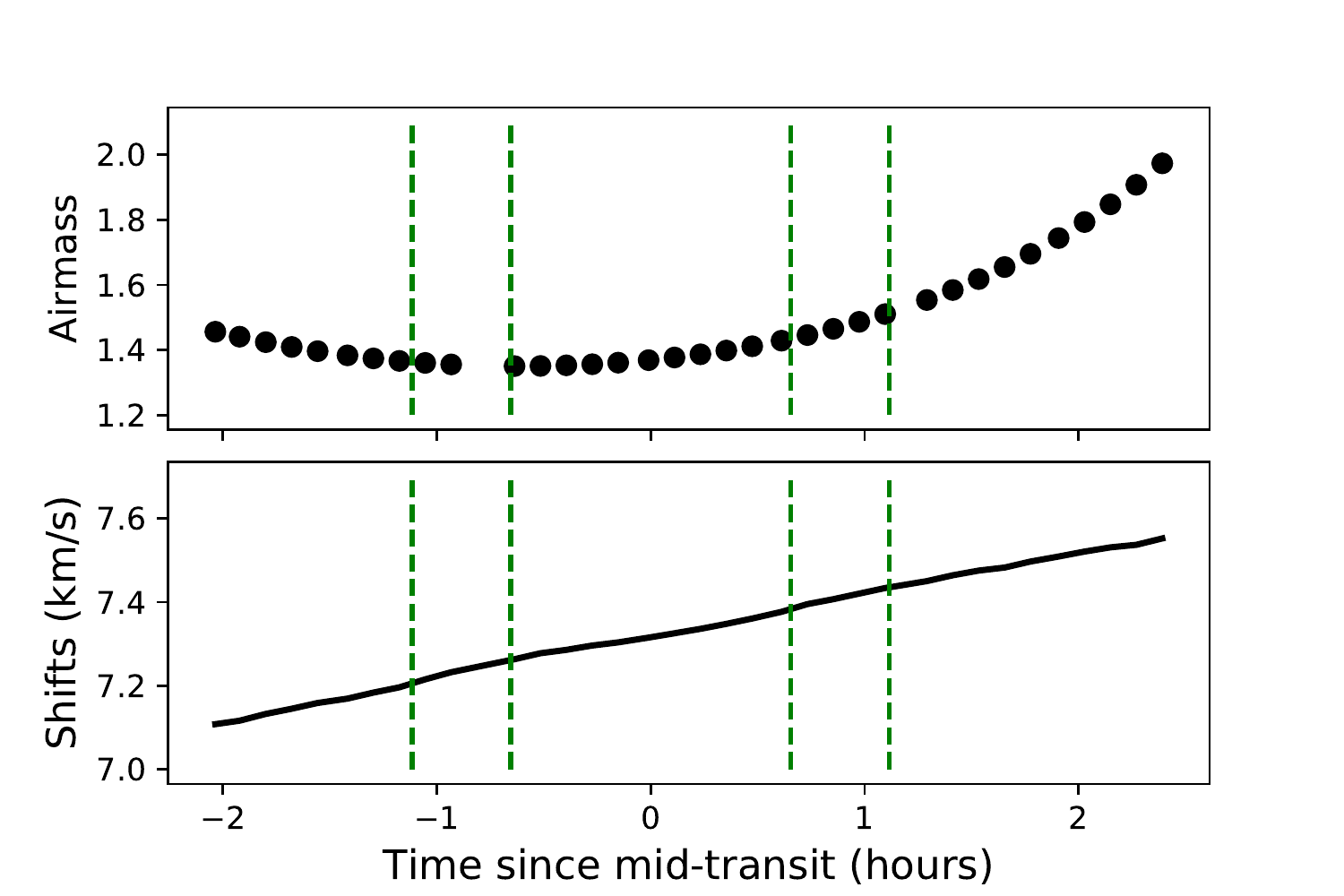}
    \caption{Airmass ({\em top}) and sum of the radial (stellar) and barycentric velocities ({\em bottom}) for each exposure during the first observing night with CARMENES. In this analysis, we discarded seven exposures with airmass higher than 1.6 to reduce the effects of contamination by the telluric features. The dashed lines represent the times of first, second, third, and fourth contact. 
    This figure can be compared with Fig.~\ref{fig:am_shift_appen}. }
        \label{fig:airmass_and_alignment}
\end{figure}

\begin{figure*}
    \centering
    \includegraphics[width = \columnwidth]{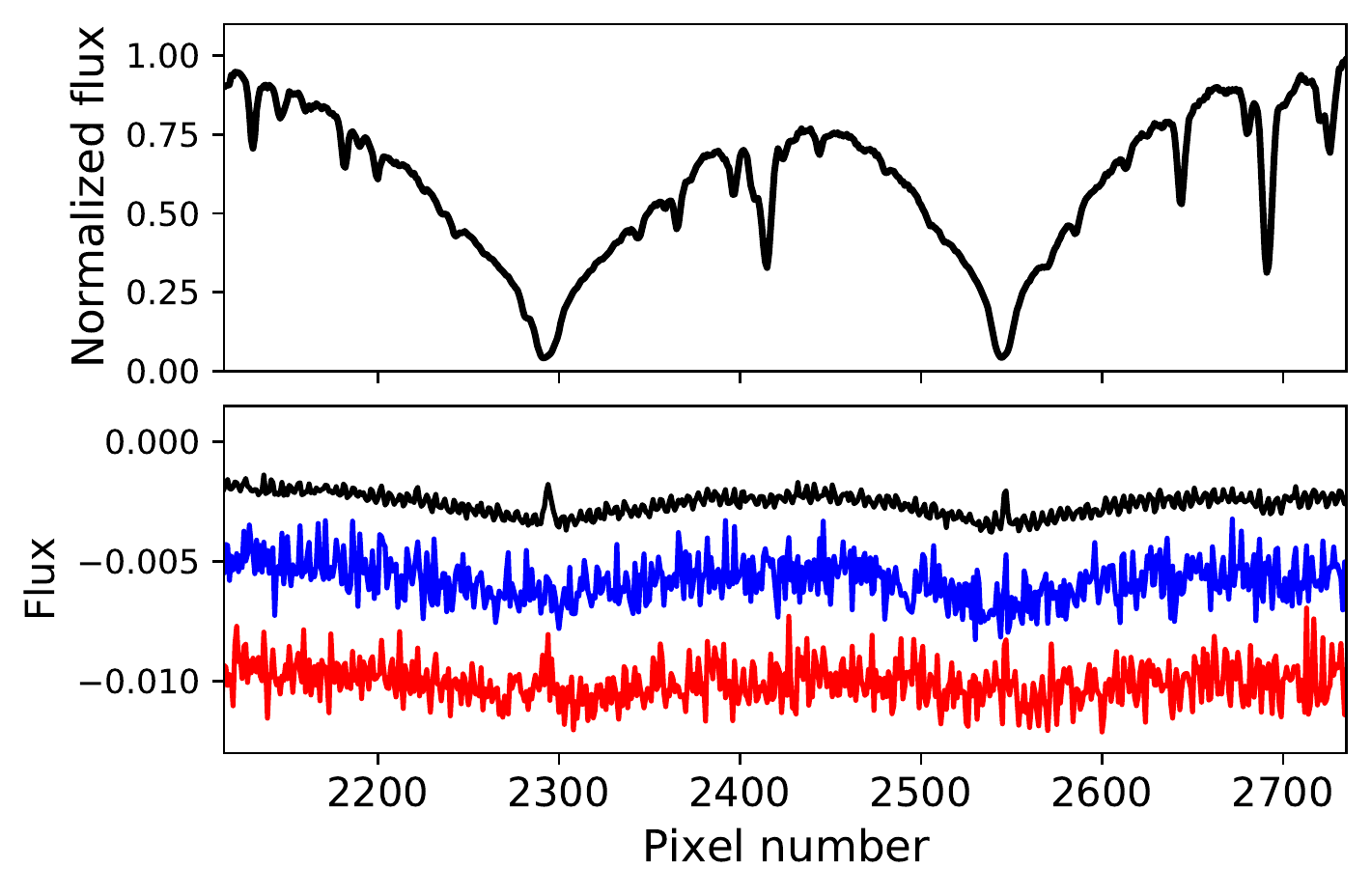}
    \includegraphics[width = \columnwidth]{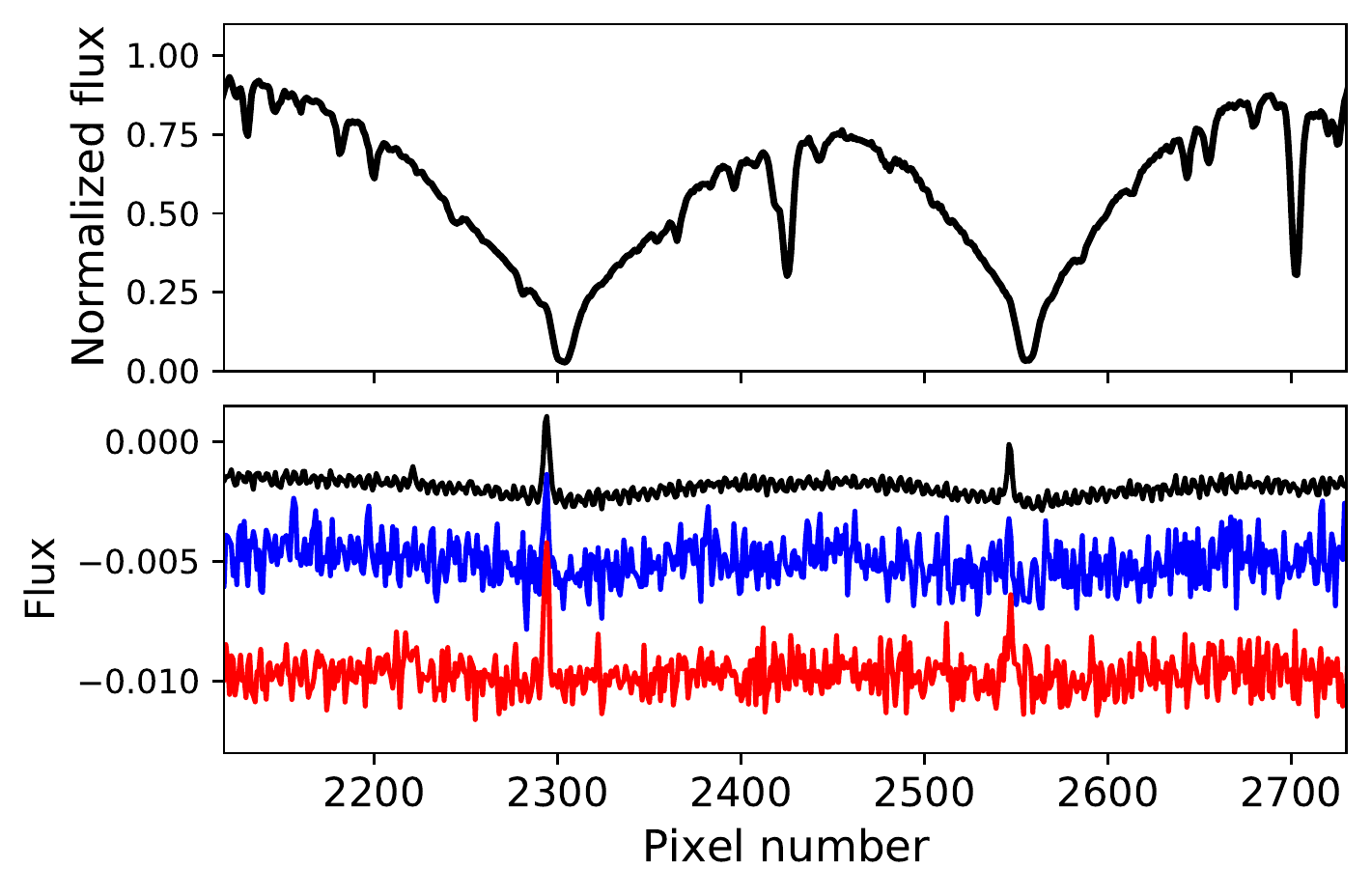}
    \caption{\textit{Left panel, top:} Average CARMENES stellar spectrum on the first night around the Na~{\sc i} D$_1$ and D$_2$ doublet.
    \textit{Left panel, bottom:} Average of all sky spectra (uppermost spectrum in black), and the lowest and highest airmass points, respectively (second (blue) and third (red) spectra, respectively).
    To avoid any offset between the wavelength solutions in fibers A and B, we use the pixel grid instead of wavelength for the x-axis. The profiles are shifted by an arbitrary offset for clarity.
    \textit{Right panels:} Same as left panels but for the second night.}
    \label{fig:FiberB}
\end{figure*}

Three transits of WASP-69~b starting on 2017 August 22, 2017 September 22, and 2020 August 13 were observed with CARMENES \citep[Calar Alto high-Resolution search for M dwarfs  with Exoearths with near-infrared and optical Echelle Spectrographs;][]{Quirrenbach2014, Quirrenbach2018} in the two visible (VIS) and near-infrared (NIR) channels. 
The details of the first two observations and the data reduction are explained by  \citet{Nortmann2018}. 
The third night also followed the same observing strategy. 
        
Below, we first summarize the information on the observations and the initial data reduction, and then explain our own data analysis of the CARMENES VIS data  in detail.
The target star was placed in fiber A, while the sky was monitored with fiber B at 88\,arcsec to the west. 
The observation from the first night consisted of 35 spectra: 18 taken before and after the transit (out-of-transit)\ and 17 during the transit (in-transit). 
On the second night, we obtained 31 spectra, consisting of 14 out-of-transit and 17 in-transit exposures. 
Finally, on the third night we captured 34 exposures, of which 17  were out-of-transit and 17 were in-transit. 
The seeing averages on the first, second, and third nights were 0.63, 0.69, and 1.11\,arcsec, respectively.
The exposure time for each frame was 398\,s in all cases, which was short enough to keep the change in planetary velocity to less than 1\,km\,s$^{-1}$. 
We used a mean CARMENES pixel equivalent-velocity size of 1.3\,km\,s$^{-1}$  so as to not smear the signal.
The airmass change on the first observing night is plotted in the top panel of Fig.~\ref{fig:airmass_and_alignment} (airmass changes on the other two nights are shown in Fig.~\ref{fig:am_shift_appen} in Appendix~\ref{sec:appen1}).
The initial raw data frames were processed through the CARMENES pipeline, {\tt caracal} \citep[CARMENES Reduction And Calibration; cf.][]{Caballero2016}. 
The spectra were aligned with the radial velocities measured by {\tt serval} \citep{Zechmeister2014}; the barycentric velocities, and the systemic velocity $\gamma$ (Table~\ref{tbl:orbital parameters}) were additionally taken into the account. 
The values of shifts from these two sources on the first night are shown in the bottom panel of Fig.~\ref{fig:airmass_and_alignment} (see also Fig.~\ref{fig:am_shift_appen} for the second and third nights). 
The average S/N of the VIS spectra are 52, 40, and 37 for the first, second, and third night, respectively.

Absorption features of the Earth's atmosphere were corrected using telluric models generated with {\tt Molecfit} \citep{Smette2015, Kausch2015}. 
In addition, to investigate the telluric emission features superimposed on the Na~{\sc i} D$_1$ and D$_2$ doublet, we used sky spectra collected simultaneously with the CARMENES fiber B. 
Figure \ref{fig:FiberB} illustrates the sky emission spectra, which shows telluric sodium emission lines right in the center of the stellar sodium lines in the average of all spectra.

These emission features are airmass-dependent and can therefore be related to the Earth's atmospheric sodium \citep{Chapman1939, Kolb1976, Noll2012} or they can be contamination by low-pressure sodium street lamps (The lights from the nearby cities of Granada and Almer\'ia are visible from Calar Alto and are sometimes reflected by high, thin cirri).
The telluric emission on the second observing night was a few times stronger than that on the first night. 
In addition, there was an instrumental contamination of flux in fiber B due to cross-talk with fiber A, which appeared in the form of a scaled spectrum of fiber A on the spectrum of fiber B. This contamination can be seen in the sky emission spectra in the lower panels of Fig.~\ref{fig:FiberB} (the continuum is not straight and is similar in shape to the stellar spectral profile).
Therefore, we were not able to simply remove the sky emission features by subtracting each spectrum with the corresponding fiber B spectrum. 
Instead, we first ignored seven spectra with airmass larger than 1.6, and then masked the emission feature as explained in Sect.~\ref{sec:2Dmap}. We conclude that the sky emission effect in the first night (after removal of those with airmass larger than 1.6) is negligible, because our final outcomes, before and after masking, are the same.

To normalize the spectra we use a method similar to that used by \citet{Khalafinejad2017}.
We first computed an averaged spectrum, and then, to obtain the continuum profile, we found the maximum data points in wavelength bins of $\sim$10\,{\AA,} ignoring the region around the targeted line (e.g., for Na D lines from $\sim$5887\,{\AA} to 5902\,{\AA} and for H${\alpha}$ from $\sim$6563\,{\AA} to 6567\,{\AA}).
To correct for outliers or any remaining spike, we substituted any point deviating by more than 2$\sigma$ with the average of its neighboring points.
We then linearly interpolated the maximum points to construct the shape of the continuum and divided each individual spectrum by this interpolated curve to flatten the averaged spectrum. 
After this stage, we again  normalized each spectrum by dividing each individual spectrum by the median of the points in the continuum.
We performed the analysis for all nights in parallel.
However, the second night was highly contaminated by telluric features (see above), and the S/Ns of the second and third nights were lower than that from the first night.
We therefore dropped the second and third nights from the final modeling stage, although the results from these nights are shown in Appendix~\ref{sec:appen1} for comparison.

\subsection{Long-term photometry with STELLA}
\label{sec:long-term}

\begin{figure}
    \centering
    \includegraphics[width=\columnwidth]{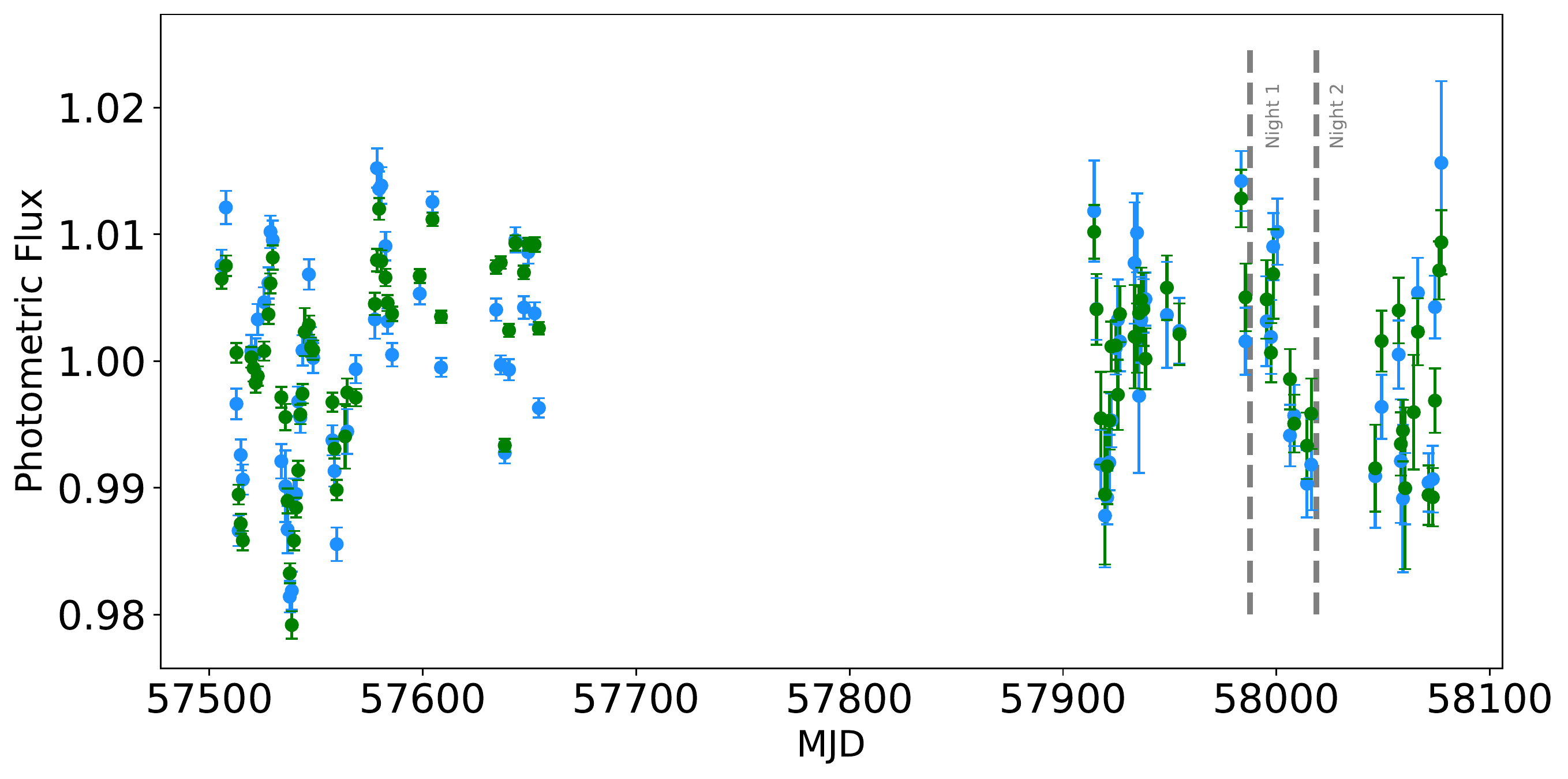}
    \caption{Normalized flux of WASP-69 in $B$ (blue) and $V$ (green) filters from our long-term photometric observations with STELLA. 
    The dashed lines mark our two first CARMENES observing nights.}
    \label{fig:photometry_stella}
\end{figure}

The host star WASP-69 was found to show signs of stellar activity by
\citet{Anderson2014}. 
The presence of spots in the stellar photosphere causes periodic variations when they rotate in and out of view. 
The stellar rotation period was estimated to be about 23\,d and the photometric variation showed a semi-amplitude of about 1\,\% \citep{Anderson2014}. 
To characterize the presence of spots on the visible stellar hemisphere during our spectroscopic transit observations, we initiated a long-term photometric monitoring campaign covering the first and second transit observations. 

We employed the robotic 1.2\,m STELLA telescope \citep{Strassmeier2004} located at the Observatorio del Teide in Tenerife (Spain), and its wide-field imager WiFSIP \citep{Weber2012}. Observations were taken from May to September 2016 (53 nights) and from June to November 2017
(40 nights) in nightly blocks of three 8\,s exposures in Johnson $B$ and three 6\,s exposures in Johnson $V$. 
The data were reduced and the light curves extracted following the procedure described by \citet{Mallonn2018}. 
In short, the image frames were trimmed, and bias- and flat-field corrected with the standard STELLA pipeline. 
We then performed aperture photometry using {\sc Source Extractor} \citep{Bertin1996}. 
We tested different aperture shapes (circular versus flexible elliptical) and different aperture sizes to finally employ the aperture minimizing the dispersion in the light curves. 
We built an artificial comparison star for target differential magnitudes as the flux sum of multiple individual sources. 
The final choice was again made by a minimization of the light-curve scatter. 
We verified that the comparison star selection had no significant influence on the photometric variability signal of WASP-69 by manually testing different sets of comparison stars. 

In Fig. \ref{fig:photometry_stella} we show the final light curve after binning the three exposures per filter of an individual observing block. 
During the second night the star appeared dimmer compared to the first night. 
In other words, the spot filling factor on the visible stellar hemisphere was smaller on night 1 than on night 2. 
One could argue that the first night of observation was obtained during a period of maximum brightness of the star, which implies that the spots were at minimum.
Unfortunately, we were not able to carry out any photometric observations simultaneous to the third CARMENES observing night in 2020.

Additionally, we re-measured the stellar rotation period.
For that, we employed the generalized Lomb-Scargle (GLS) periodogram \citep{Zechmeister2009} and the corresponding false-alarm probabilities (FAPs) to search for significant peaks from the available photometric $B$- and $V$-band light curves from both 2016 and 2017. 
We obtained peaks at 24.34\,d and 24.95\,d above the 0.1\,\% FAP threshold for the 2016 $B$- and $V$-band data sets, respectively. 
We interpreted the $\sim$24\,d signal as the stellar rotation period, as it is very close to the 23.07\,$\pm$\,0.16\,d period reported by \citet{Anderson2014}. As a reference, periodograms and phase-folded light curves related to this observation are shown in Fig. \ref{fig:per-gram}.
In comparison, for the 2017 data sets, the GLS periodograms show peaks at 48.24\,d above the 10\,\% FAP level in $B$ band and above the 0.1\,\% FAP level at 11.70\,d in $V$ data, which could be considered as the first harmonics of the $\sim$24\,d signal.
The one-day difference between the measured rotation periods can probably be attributed to differential rotation.

\section{Transmission spectroscopy}
\label{sec:TS}

\begin{figure}
    \centering
    \includegraphics[width=\columnwidth]{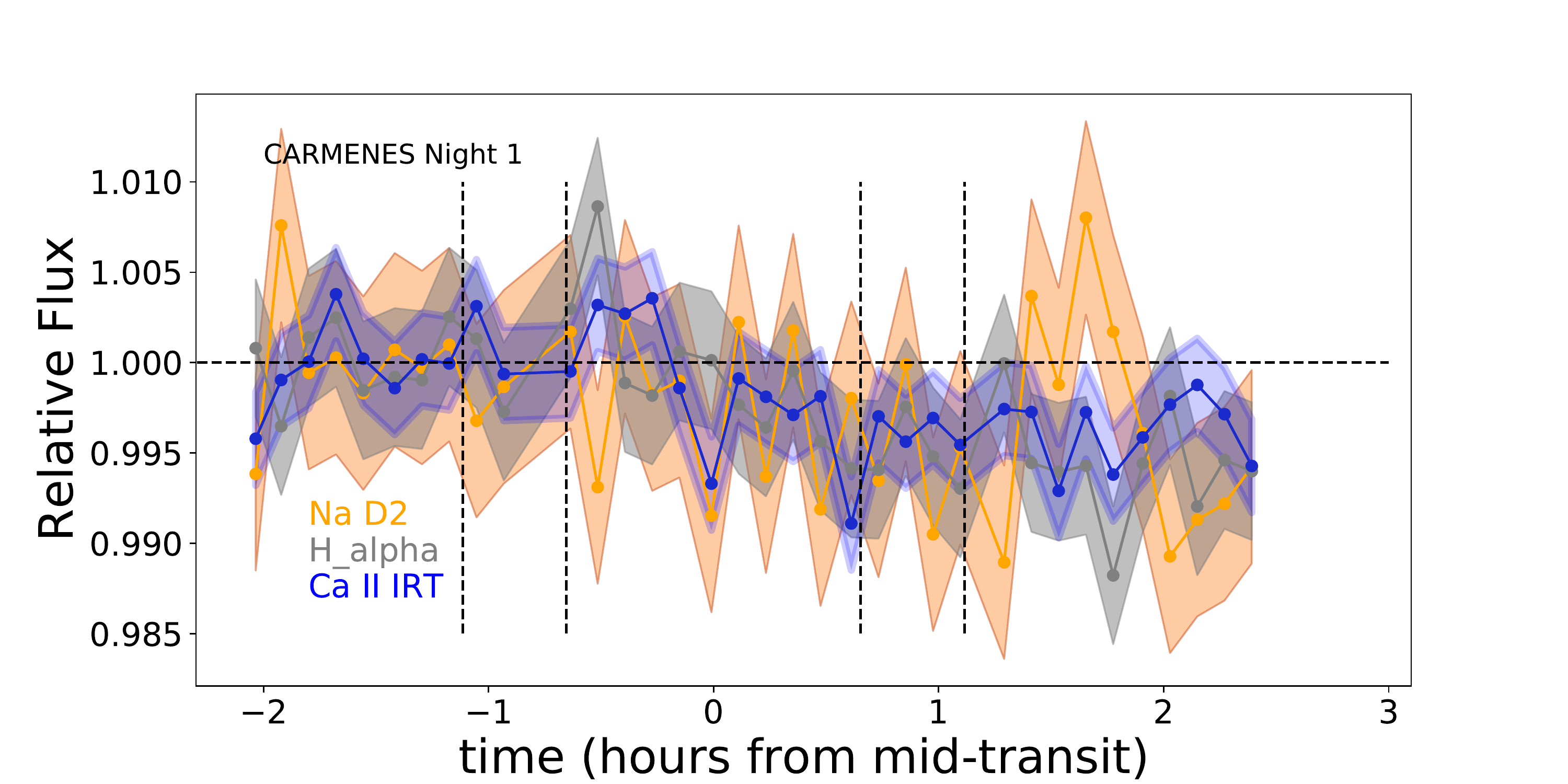}
    \includegraphics[width=\columnwidth]{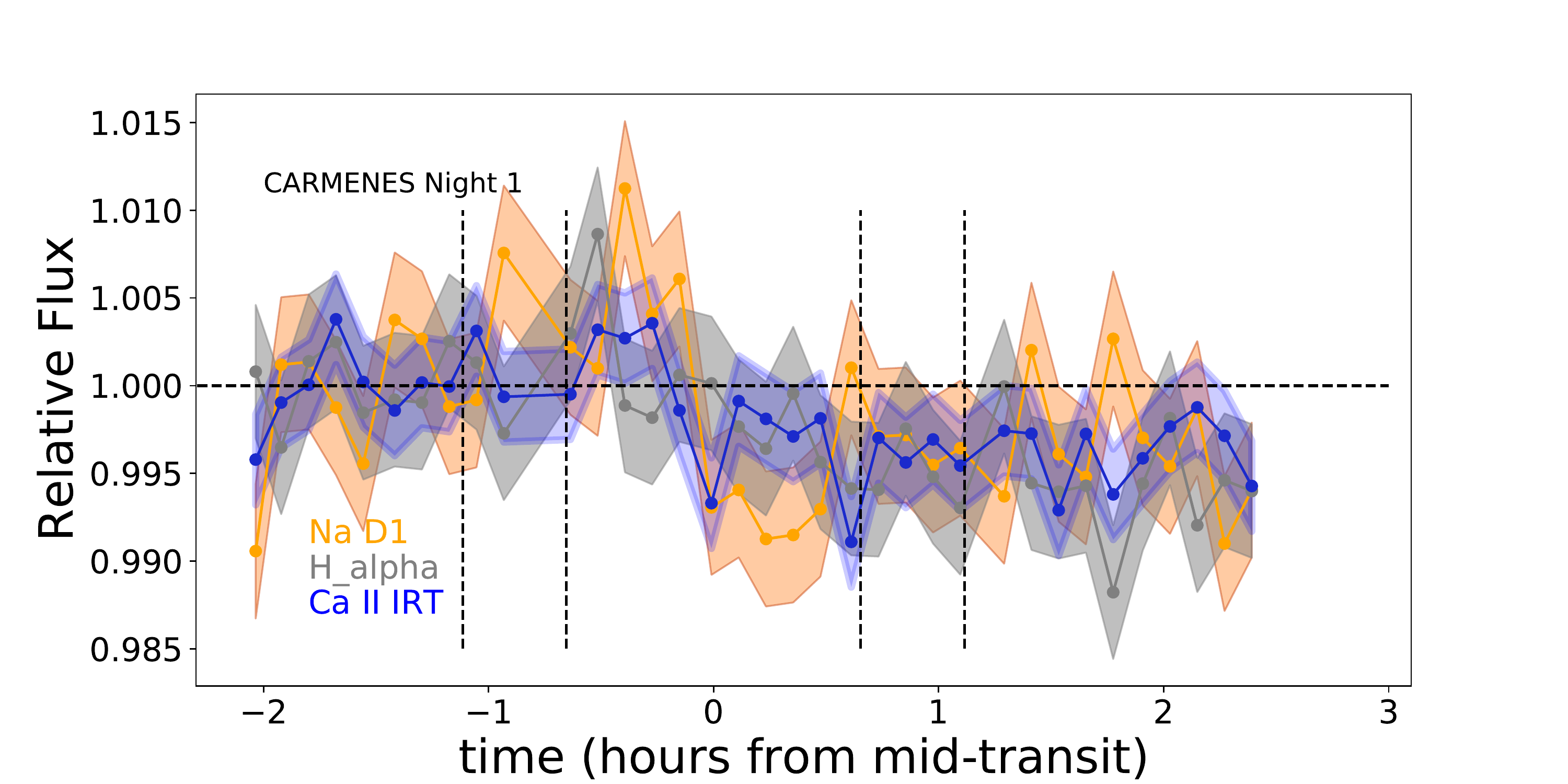}
    \caption{Relative flux time evolution of stellar line cores of \ion{Na}{i} D$_{2}$ ({\em top panel}, orange) and D$_1$ ({\em bottom panel}, orange) compared with H${\alpha}$ (gray) and the \ion{Ca}{ii}~{IRT} (blue) on night~1.
    The vertical dashed lines mark the times of first, second, third, and fourth contact, and the horizontal dashed line is unity.
    This figure can be compared with Fig.~\ref{fig:lc_appen}.}
    \label{fig:chrom_activity}
\end{figure}

In this section we describe the data analysis of the high-resolution transmission spectroscopy performed with CARMENES (for all of the three nights) around the strong lines of Na~{\sc i} D$_1$ and D$_2$, H$\alpha$, Ca~{\sc ii} infrared triplet (IRT), and K~{\sc i} $\lambda$7699\,{\AA}.
At the end of this section we discuss the results related to the detected signals.

\subsection{Stellar activity and excess light curves}
\label{sec:stellar_activity}

\begin{figure*}
    \centering
    \includegraphics[width=\textwidth]{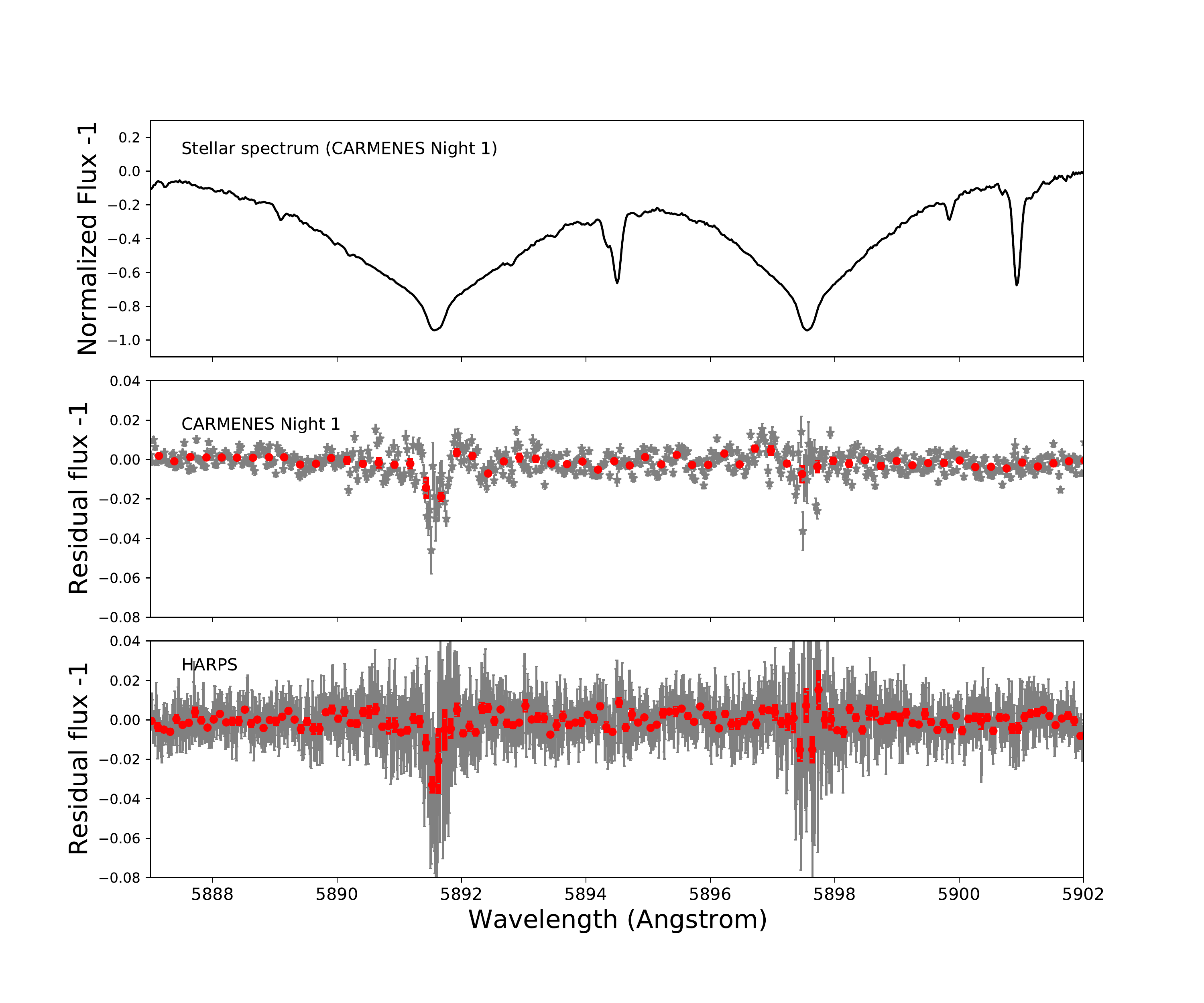}
    \caption{\textit{Top panel:} Stellar spectrum (telluric absorption corrected) of WASP-69 around the Na~{\sc i} doublet. 
    \textit{Middle panel:} Planet transmission spectrum of WASP-69~b in the same wavelength range. 
    The gray data points show the division of in-transit by out-of-transit data, corrected for the changing orbital velocity of the planet.
    The red filled circles are the data binned by 0.25\,{\AA} and the red error bars are the standard deviation of each group of binned data points.
    \textit{Bottom panel:} Planet transmission spectrum obtained with HARPS-N by \citet{Casasayas-Barris2017}, shown for comparison. 
    Both CARMENES and HARPS-N data show a similar pattern.}
    \label{fig:TS_Na}
\end{figure*}

The time evolution of the equivalent widths of some prominent stellar lines (e.g., \ion{Ca}{ii} {H\&K}, H${\alpha}$, \ion{Ca}{ii} {IRT}) is an indication of stellar chromospheric activity \citep[e.g.,][]{Linsky1979, Notsu2013, Klocova2017}. 
To investigate the possible influences of the stellar activity on the exoplanetary signals, we first investigated the temporal evolution of the core of the H$\alpha$, \ion{Ca}{ii} {IRT}, and Na D lines, and compared them to each other. 
In this section, we use the term ``light curve'' to refer to the time-series of the line cores (line indices).
For this purpose, we first look for the center of each line by fitting a Gaussian, integrating the flux in a 1.5\,{\AA} pass-band centered on the core of the line, and dividing it by the integrated flux of the reference band in the continuum \citep{Khalafinejad2018}. In principle, a Lorenzian profile can better explain the line shape compared to a Gaussian profile; however, due to the relatively large scatter of the data here, neither one has an advantage over the other.
We normalized the light curves by dividing the data points by the median of the out-of-transit level. 
We estimated the uncertainty with the scatter of the out-of-transit data. 
The time-series obtained by this method are known as ``excess light curves'' and they have the same pattern as the time evolution of the equivalent width of the lines. 
Figure~\ref{fig:chrom_activity} shows the corresponding excess light curves for the first night for the H${\alpha}$, \ion{Ca}{ii} {IRT}, and Na D lines individually. 
Excess light curves for nights 2 and 3 can be seen in Fig. \ref{fig:lc_appen}. The star did not show any strong chromospheric flare during any of the nights. The excess light curves for the \ion{Ca}{ii} {IRT} together with He~D$_3$ were also shown by \citet{Nortmann2018} for the first and the second nights. Our re-analysis of the \ion{Ca}{ii}~{IRT} in these nights agrees with their result.

However, the excess light curves display a mild level of stellar activity that reached a maximum value of about 1\,\% in the cores of H${\alpha}$ and the \ion{Ca}{ii} {IRT}. 
On night 1 (Fig.~\ref{fig:chrom_activity}), the Na D lines behaved similarly to H${\alpha}$ and the \ion{Ca}{ii} {IRT}. 
On nights 2 and 3, as shown in Fig.~\ref{fig:lc_appen}, the Na D lines showed larger fluctuations compared to H${\alpha}$ and the \ion{Ca}{ii} {IRT}. 
One possible explanation for these fluctuations is a lower S/N at 5892--5898\,{\AA} compared to the other lines at redder wavelengths.
In general, the excess light curves do not show any clear atmospheric absorption trend, and so we used other approaches for further investigation of the exoplanetary atmosphere.

\subsection{Transmission spectroscopy at Na~{\sc i} D$_1$ and D$_2$}
\label{sec:TS_Na}

\subsubsection{Division approach at Na D}
\label{sec:TS_Na_division}
\begin{figure*}
    \centering
    \includegraphics[width=0.49\textwidth]{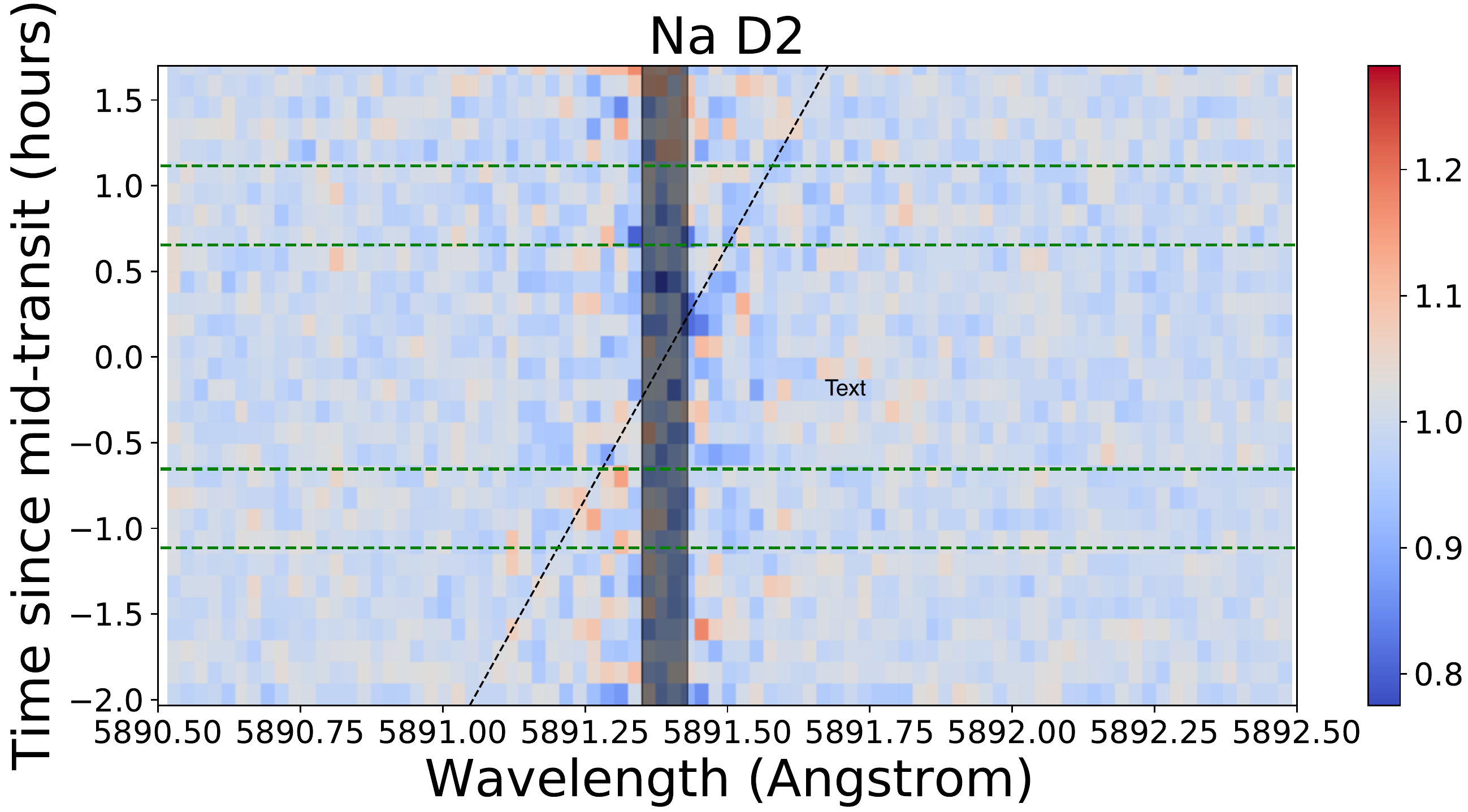}
    \includegraphics[width=0.49\textwidth]{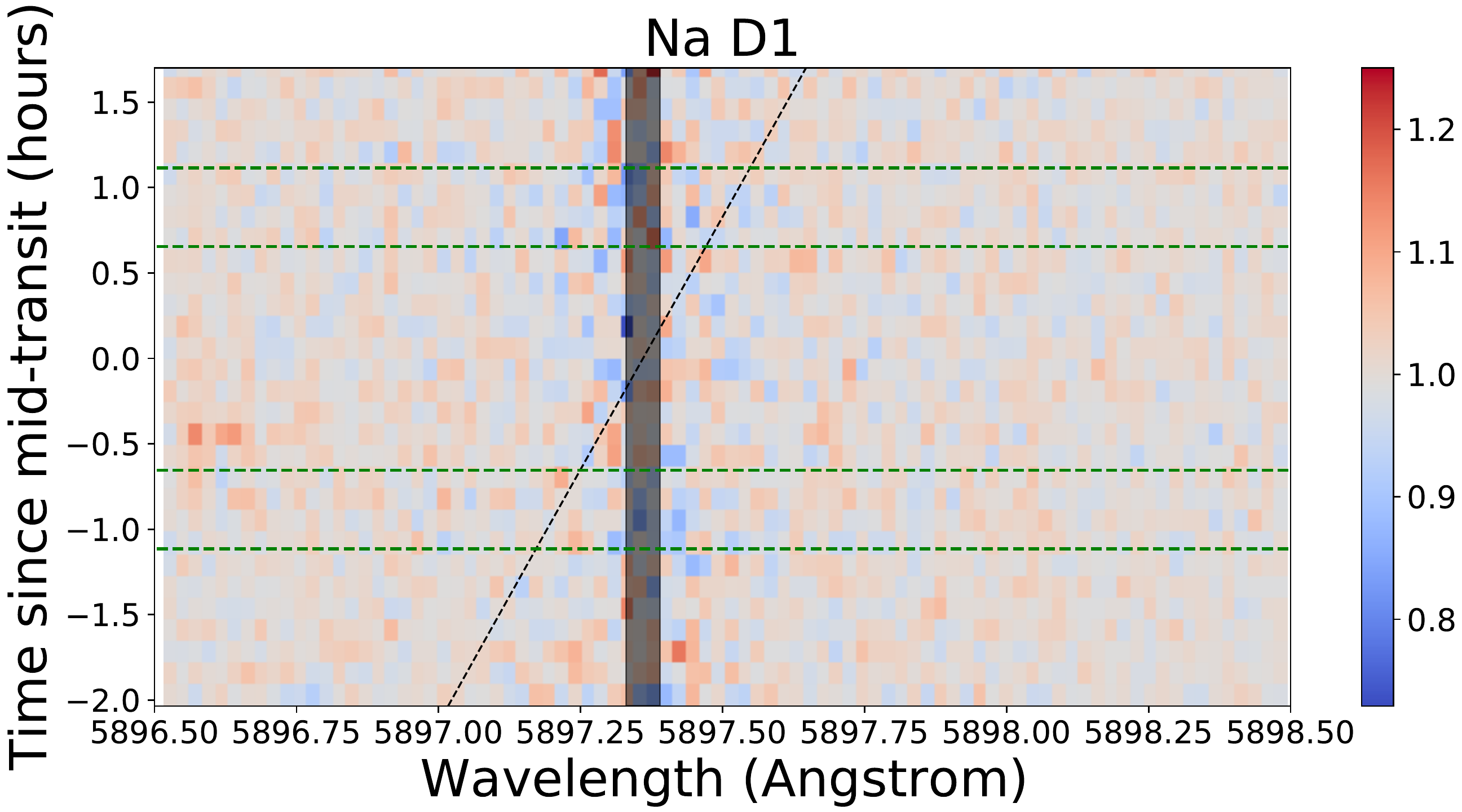}
    \caption{Two-dimensional map of residuals for both Na~{\sc i} D$_{2}$ ({\em left}) and D$_{1}$ ({\em right}) lines for the first night. 
    The green horizontal dashed lines show the times of first, second, third, and
fourth contact and the diagonal dotted line represents the expected trail of the exoplanetary atmospheric absorption.
    In each panel, the vertical axis is time (similar to the exposure number). 
    Blue and red colors stand for residual fluctuations in the direction of absorption and emission, respectively. The tested masked region is under the shaded area.
    } 
    \label{fig:matrix_plot}
\end{figure*}

To obtain the transmission spectrum using high-resolution data, a simple method is to divide the average of in-transit spectra by the average of out-of-transit spectra, called ``master-out''. 
However, because the planet is moving around the star, this approach must be corrected by taking into account the radial velocity of the planet. 
We obtained the transmission spectrum of WASP-69~b using the method described in detail by \citet{Khalafinejad2018}. Briefly, we divided each in-transit spectrum by the master-out and then aligned the residual spectra based on the radial velocity of the exoplanet. At the end, we co-added all residual spectra, computed the average, and considered it as the exoplanetary transmission spectrum  for each transit observation
(Figs.~\ref{fig:TS_Na} and \ref{fig:TS_Na_app}). 
In the transmission spectrum of the first night, shown in the middle panel of Fig.~\ref{fig:TS_Na}, the core of the Na~{\sc i} D$_{2}$ line shows an absorption at the level of about 2\,\%. 
However, Na~{\sc i} D$_{1}$ does not show a signature of any considerable absorption. 
For comparison, we also show the stellar spectrum in the first panel, and the transmission spectrum by \citet{Casasayas-Barris2017} from HARPS-N observations in the bottom panel of Fig.~\ref{fig:TS_Na}. 
The results from the first night of observation with CARMENES agree well with those from the HARPS-N data.
The error bars on each data point were obtained by propagating the errors associated with the raw spectra. 
For the investigation of the physical properties of the atmosphere, we merged the transmission spectrum of the first CARMENES night with that of the HARPS-N data (see Sect.~\ref{sec:modeling}). As mentioned in Sect.~\ref{sec:obs_carmenes}, the second night was highly affected by the telluric features and in the third night the S/N was relatively low. Finally, the transmission spectrum of the second night shows large scatter with no sign of exoplanetary absorption. The results from the third night show possible signs of exoplanetary absorption, but with more scatter than the first night. Hence, we exclude these nights from our atmospheric modeling in Sect.~\ref{sec:high-res}. 

\subsubsection{Two-dimensional map of residuals at Na D}
\label{sec:2Dmap}

Another method for visualizing the absorption features of exoplanetary atmospheres is to investigate the matrix of residual spectra obtained from the division of each individual spectrum by the master-out stellar spectrum. 
In Fig.~\ref{fig:matrix_plot}, we show the map of residual spectra at each sodium line for the first night of CARMENES observations (see Fig.~\ref{fig:matrix_plot_app} for the second and third nights). 
The expected exoplanetary signatures are those beneath the dotted diagonal lines.
The vertical feature in the background in each panel is the lower S/N region originating from the core of the stellar sodium lines. 
We did not see a clear trail of exoplanetary Na D absorption in these maps. 
The lowest pixels could be due to the telluric emission or the variability of stellar emission core and we needed to make sure that the entire signal in the transmission spectrum (obtained through the division approach, see Sect.~\ref{sec:TS_Na_division}) was not based on only these few ``potential outliers''. Hence, we removed the exposures containing very low pixels on the expected exoplanetary trail and repeated the analysis.
We observed no significant changes in the shape of the transmission spectra after this procedure. 
In addition, we completely masked the narrow band contaminated by the telluric emissions based on Fiber B spectra. The masked region is shown under the shaded area in Fig.~\ref{fig:matrix_plot}. Masking the whole region did not change the general shape of the transmission spectrum or the absorption level; it simply led to larger error bars in the transmission spectrum for the points in the core of the line, because fewer residual spectra formed this region. 
This confirmed that the sky emission effect (after removal of the seven last spectra with the highest airmass) was within the error bars of the signal. At the end, we did not mask the region to keep the highest possible number of in-transit spectra.

Considering that the 2D map does not reveal any exoplaneary signature, the best way to detect a signal is to co-add the signal in all in-transit exposures, as explained in Sect.~\ref{sec:TS_Na_division}.

\subsection{Transmission spectrum at H${\alpha}$ and other lines}
\label{sec:TS_halpha}

\begin{figure*}
    \centering
    \includegraphics[width=0.49\textwidth]{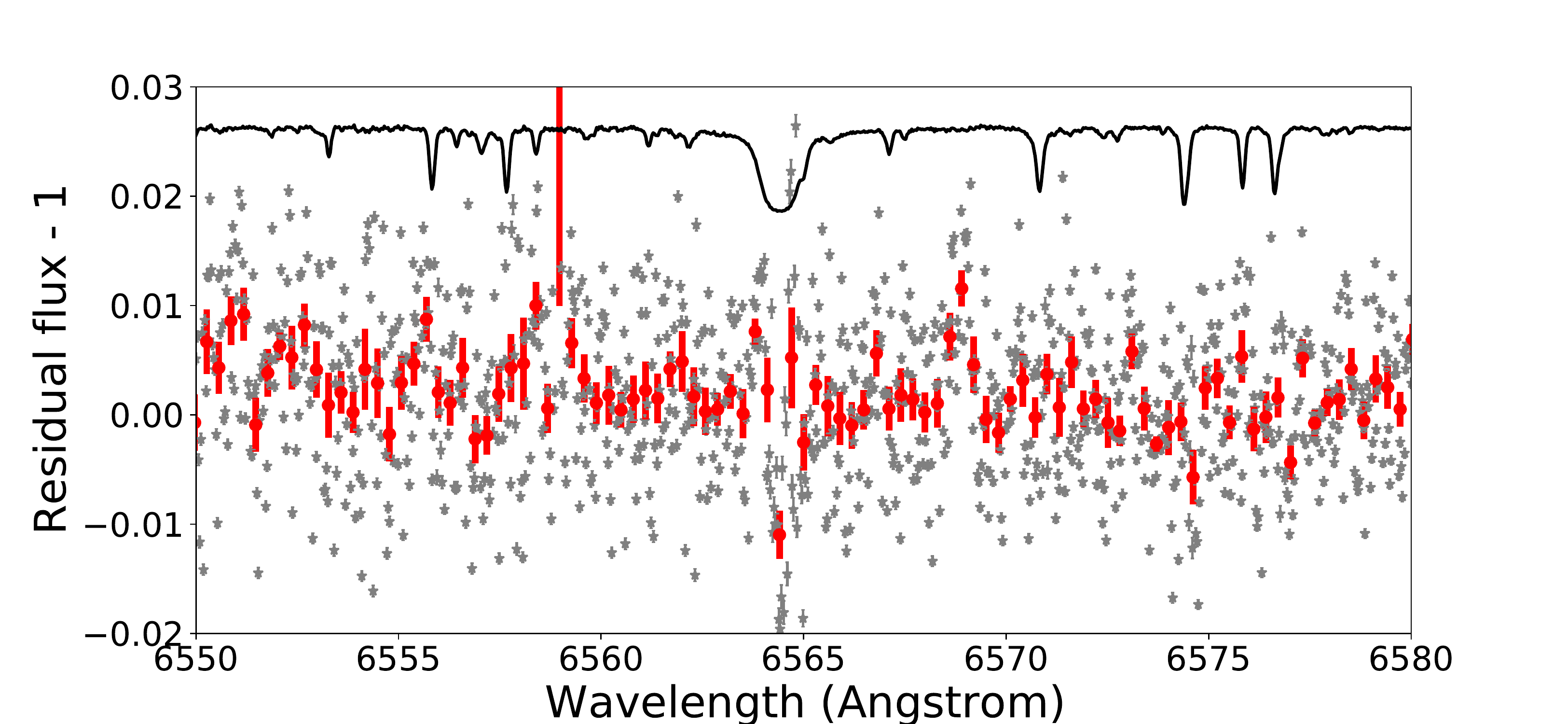}
    \includegraphics[width=0.49\textwidth]{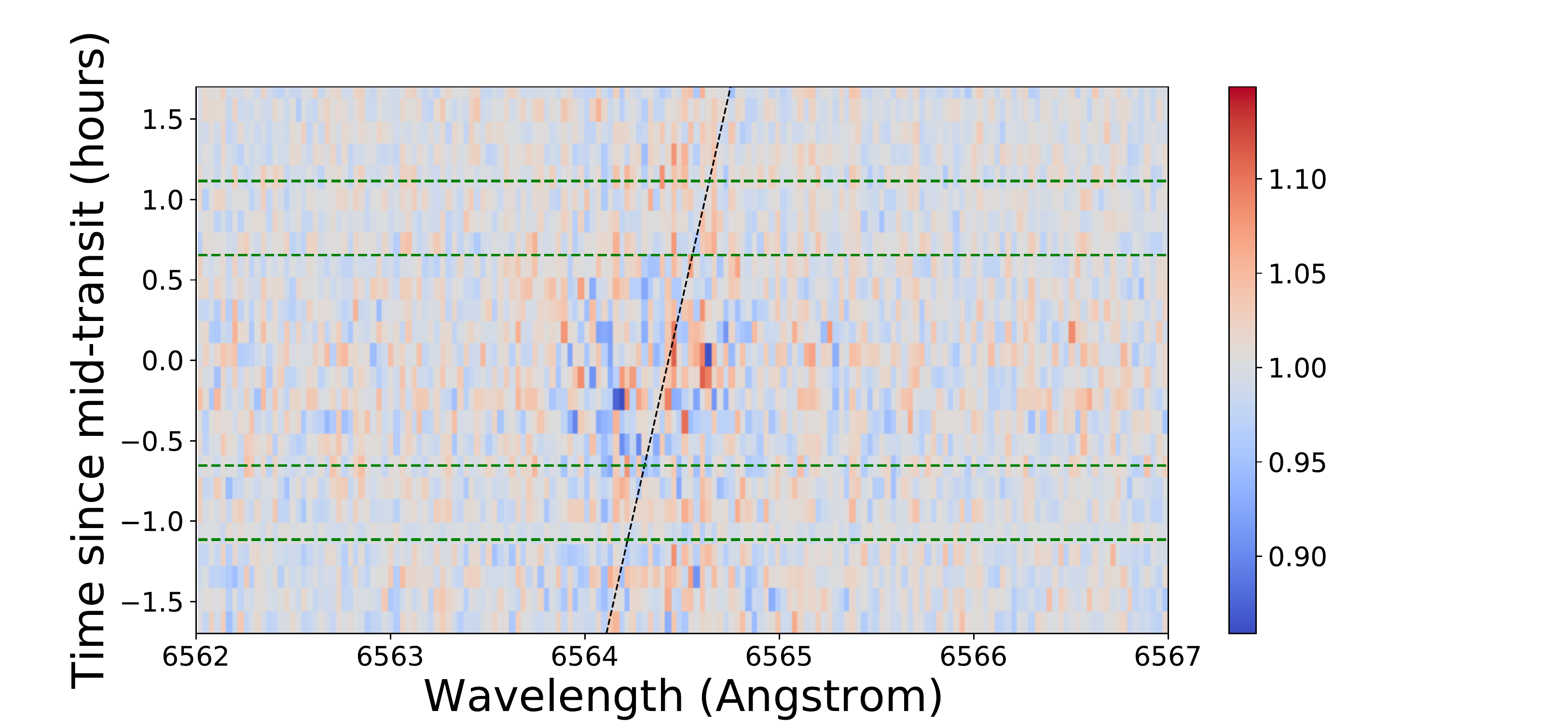}
    \caption{{\em Left}: Transmission spectrum around H${\alpha}$ for the third night. 
    The black line is the scaled and shifted stellar spectrum, the gray dots with gray error bars are the residuals of the division, and the red filled circles with red error bars are the data binned by 0.3\,\AA\, (here, the error bars are the standard deviation of each group of binned data points).
    {\em Right}: Same as Fig.~\ref{fig:matrix_plot} but for H${\alpha}$.}
    \label{fig:TS_matrix_plot_halpha}
\end{figure*}

\begin{figure*}
    \centering
    \includegraphics[width= 0.49\textwidth]{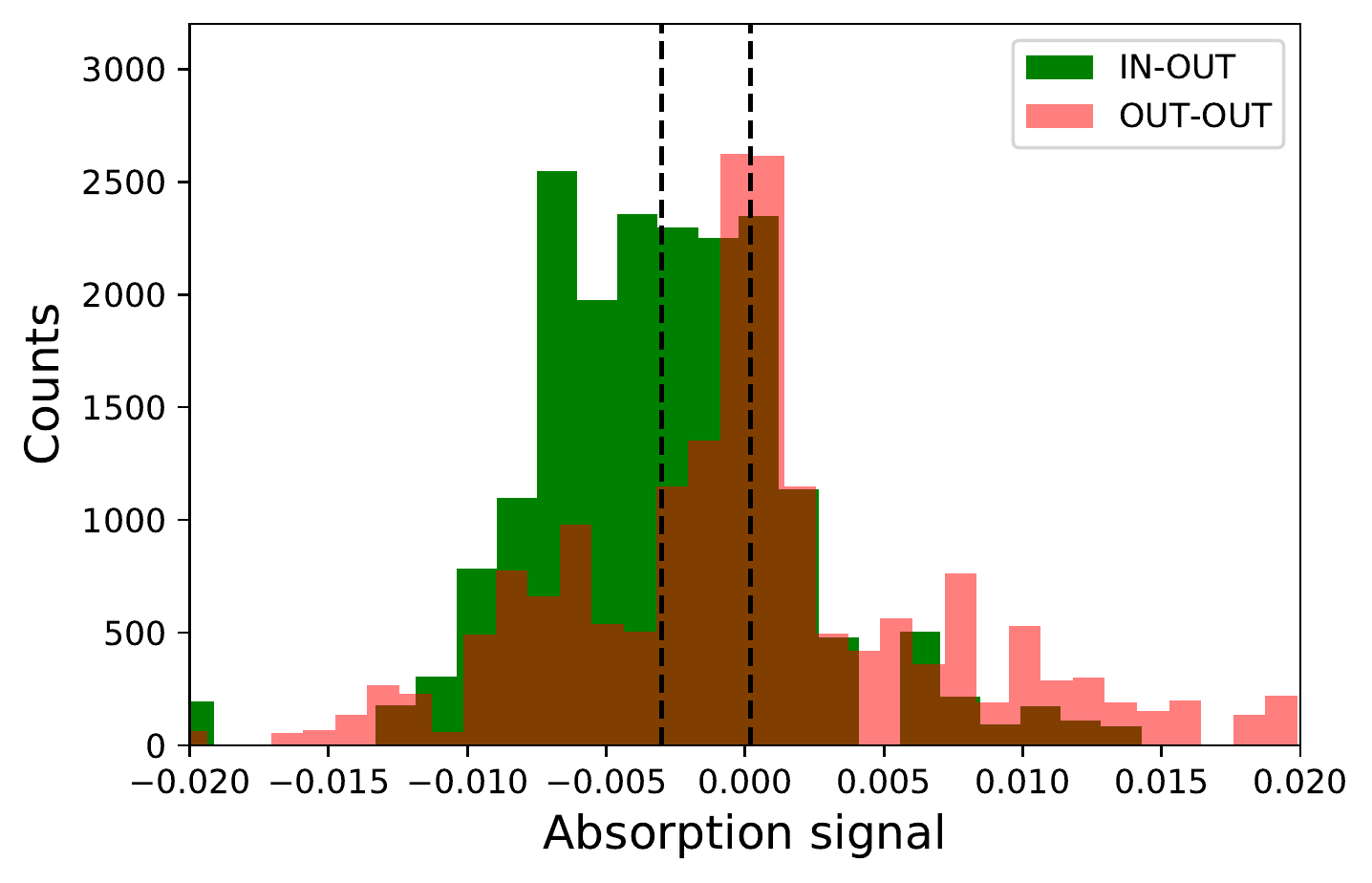}
    \includegraphics[width= 0.49\textwidth]{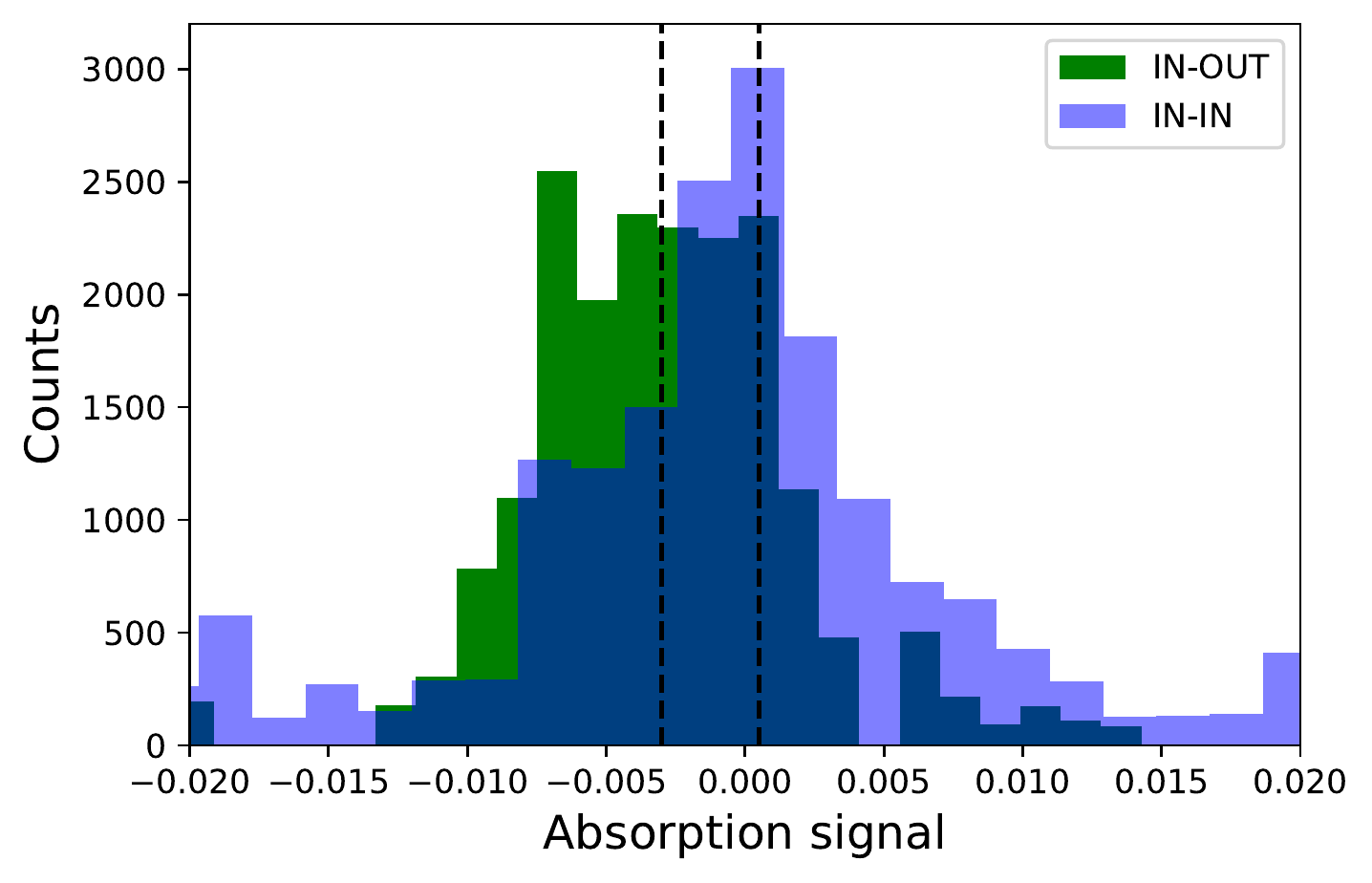}
    \caption{Distributions of the empirical Monte Carlo bootstrap analysis of the core of Na D features on the first night. 
    The ``in-in'' (blue) and ``out-out'' (red) distributions are centered around zero (no sodium detection), but the randomized in-out (green) distribution shows a detection. 
    The dashed lines show the center of the Gaussian fit over each distribution.}
    \label{fig:hist}
\end{figure*}

For all nights, we investigated the exoplanetary H${\alpha}$ line using the division approach and the 2D map of residuals, as for the Na D lines. 
We found no signatures of H${\alpha}$ except for the third night, which displayed a weak signature (Fig.~\ref{fig:TS_matrix_plot_halpha}).
Both the transmission spectrum and the matrix plots show hints of signatures of H${\alpha}$ absorption for that night. 
The 2D map shows a relatively uniformly distributed clump of low pixels on the expected exoplanetary trail, which is similar to the He~{\sc i} $\lambda$10830\,{\AA} absorption feature 
shown by \citet[][their Fig. S2]{Nortmann2018}.
In the transmission spectrum, there is also a signature of absorption at H${\alpha}$.
However, we refrain from claiming a detection because fluctuations at a similar level are visible in the vicinity of this line. 
In addition, no feature was detected at the H${\alpha}$ line in the other nights, particularly in the first night that had the highest S/N.


With a method similar to the analysis of the Na D and H${\alpha}$ lines, we investigated the Ca~{\sc ii} infrared triplet (IRT) and K~{\sc i} $\lambda$7699\,{\AA} lines.
Using the division approach and the 2D representation of the residual spectra in the vicinity of these lines, we observed no sign of absorption for any of the lines.

\subsection{Signal strength at Na D}
\label{sec:discuss_Na}

In order to compare our results with previous studies, we applied a Gaussian toy model to the Na D doublet region of our transmission spectrum to estimate the strength of the absorption signal. 
We used the sum of two Gaussian profiles centered on the lines of the Na D doublet and applied a Markov chain Monte Carlo (MCMC) procedure using {\tt emcee} \citep{Foreman2013} to determine the best model parameters and their uncertainties. The details of this procedure including the prior and posterior values are described in Appendix \ref{sec:appen_Gauss}.

We measured absorption signals of 0.7$\pm$0.1\% at Na~{\sc i} D$_{2}$ and 0.3$\pm$0.1\% at Na~{\sc i}  D$_{1}$, both in a passband of 1.5\,\AA. 
In other words, we detected a 7$\sigma$-level excess absorption at Na~{\sc i} D$_{2}$ and a 3$\sigma$-level excess absorption at Na~{\sc i} D$_{1}$. 
For estimating the D$_{2}$/D$_{1}$ ratio, we took the product of $\sigma$ and amplitude for each line and obtained a D$_{2}$/D$_{1}$ ratio of 2.5 $\pm$ 0.7. 

In our analysis, the significance of the detection at sodium D$_{1}$ is less than for D$_{2}$. This is consistent with the results of \citet{Casasayas-Barris2017}, who do not detect an absorption at Na D$_{1}$. 
In general, the absorption cross-section
of the D$_{2}$ line is up to two times larger than in D$_{1}$ \citep{Draine2011,Gebek2020}.

In our case, the D$_{2}$/D$_{1}$ ratio is about 2.5 $\pm$ 0.7, which tends to be closer to the upper limit of what we expect.
We further investigate the physical properties of the atmosphere inferred from these lines in Sect. \ref{sec:high-res}. 
The reason for the non-detection of sodium by \citet{Deibert2019} could be related to the fact that the signal at the sodium D$_{1}$ line is relatively weak and the stronger signal from the sodium D$_{2}$ line is not enough to appear in the cross-correlation method.

We ignored the center-to-limb variation and the Rossiter-McLaughlin effect. 
Based on \citet{Casasayas-Barris2017}, the amplitude of the former is only a small fraction (about one-tenth at 1.5\,\AA) of the expected absorption depth. 
In addition, the star is not a fast rotator ($v\sin{i} =2.2 \pm 0.4$\,km\,s$^{-1}$) and the planetary orbital velocity ($\sim$127\,km\,s$^{-1}$) is much faster than stellar rotation. 
Therefore, \citet{Casasayas-Barris2017} did not find a significant Rossiter-Mclaughlin effect.

\subsection{Signal significance at Na D}
\label{sec:discuss_stat_Na}

To check the probability of a false positive in our transmission spectra, we applied a bootstrap (or empirical Monte Carlo) analysis similar to those implemented by \citet{Redfield2008} and \citet{Seidel2020}. 
We extracted subsamples from both in-transit and out-of-transit spectra and randomly made three sets of transmission spectra. 
The first group was the division of a random in-transit spectrum by a random out-of-transit spectrum (``in-out''), the second group a random division of an out-of-transit  spectrum by an out-of-transit spectrum (``out-out''), and the third group a random division of an in-transit  spectrum by an in-transit spectrum (``in-in'').

Similar to the procedure in Sect. \ref{sec:discuss_Na}, we fitted a Gaussian function to each of the division residuals and estimated the absorption signal at the D$_{2}$ line (i.e., the area under the Gaussian). For a robust detection, the generated transmission spectra in the second and third groups should show no absorption, meaning that their distribution of the absorption signal must be around zero. 
However, we expect the absorption signal distribution for the first group to be negative.

The distribution of the absorption signals after 20\,000 iterations is shown in Fig.~\ref{fig:hist}. 
By measuring the mean of these distributions, we measured the peak of the ``in-out'' absorption signal at the negative value of --0.0030$\pm$0.0001, while the other two groups were distributed around zero with the mean value of 0.0002$\pm$0.0001 for ``out-out'' and 0.0005$\pm$0.0002 for ``in-in''. 
We considered the false-positive likelihood as the standard deviation of the ``out-out'' distribution multiplied by the square root of the fraction of out-of-transit spectra of the total number of spectra to account for the sample selection \citep{Redfield2008, Astudillo2013, Wyttenbach2015, Seidel2020}.
In this case, our false-positive likelihood was 0.4\,\%, which was still small enough to conclude 
that we measured a genuine absorption signal from the planet.

\section{Atmospheric modeling}
\label{sec:modeling}

We characterized the atmosphere of WASP-69~b using a two-step retrieval approach. First, we performed a low-resolution atmospheric modeling and then we used the output parameters of the best-fit model, including the continuum level information (i.e., haze opacity, atmospheric temperature, and base pressure) as priors for our high-resolution modeling around the Na~D features.

\subsection{Low-resolution atmospheric modeling}
\label{sec:low-res}

To date, several groups have modeled the WASP-69~b atmosphere in retrieval. \citet{Tsiaras2018} and \citet{FisherHeng2018MNRAS} used the WFC3/{\em HST} observations and retrieved the planetary (isothermal) temperature, radius, water abundance, and cloud properties. Their conclusions were similar, showing a radius close to that of Jupiter, a muted water feature, the presence of non-gray aerosols, solar water abundance, and a relatively low atmospheric temperature (492$\pm$53\,K and 658$^{+148}_{-107}$\,K, respectively in each work) consistent with a high albedo. Recently, \citet{Murgas2020-WASP69b} combined these WFC3/{\em HST} data with their OSIRIS/GTC optical observations and explored Rayleigh scattering in addition to the atmospheric carbon-to-oxygen ratio (C/O) and metallicity. These authors also investigated the activity of WASP-69 and the presence of stellar spots, testing whether the observed slope seen in the data (the increase in the apparent planet-to-star radius ratio towards the blue wavelengths) could be a consequence of unocculted stellar spots instead of aerosol particles present in the planetary atmosphere \citep{PontEtal2013, McCulloughEtal2014}. Their analyses showed that the slope seen in the data is more likely caused by hazes present in the atmosphere of WASP-69~b rather than by stellar active regions, revealing a planet with strong Rayleigh scattering,  a subsolar C/O ratio, high metallicity, and a super-solar water abundance. Furthermore, they retrieved much higher temperatures than previous studies (1203$^{+123}_{-429}$\,K and 1227$^{+133}_{-164}$\,K for the cases with and without stellar spots, respectively), which are difficult to reconcile with standard models of planetary atmospheres, even assuming a cloud-free atmosphere \citep[][see also Table~\ref{tbl:orbital parameters} in Sect.~\ref{sec:intro}]{Carone2021}. 

\subsubsection{Modeling setup}
\label{sec:model}

To assess the atmospheric properties of WASP-69~b from low-resolution spectroscopy, we performed several retrieval analyses with {\tt PyratBay} \citep[Python Radiative-transfer in a Bayesian framework,][]{CubillosBlecic-PyratBayI, BlecicEtal2021-TSC, BlecicEtal2021-DRIFT} on the joint WFC3/{\em HST} \citep{Tsiaras2018} and OSIRIS/GTC \citep{Murgas2020-WASP69b} datasets. {\tt PyratBay} is an open-source retrieval framework that generates 1D atmospheric models of planetary temperatures, species abundances, altitude profiles, and cloud coverage, using up-to-date knowledge of the atmospheric physical and chemical processes and alkali, molecular, collision-induced Rayleigh and cloud opacities. The code can work in both forward and retrieval modes, utilizing self-consistency or different parameterization schemes when including atmospheric processes. For this study, we applied basic physical assumptions without exploring the hierarchy of models available in the {\tt PyratBay} framework. We generated several modeling setups for the purpose of providing the baseline information and inputs for the high-resolution analysis presented in Sect.~\ref{sec:high-res}. As part of a coordinated effort, \citet{BlecicEtal2021-WASP-69b} will present detailed forward and retrieval analyses covering different levels of model complexity and discuss their effect on the WASP-69~b spectrum. 

In our analysis, we also combined the WFC3/{\em HST} \citep{Tsiaras2018} and OSIRIS/GTC \citep{Murgas2020-WASP69b} data and explored additional models. We performed two retrievals using two different datasets. Both retrievals included the WFC3/{\em HST} data from \citet{Tsiaras2018} (denoted WFC3 in the following text), but the first included only the OSIRIS/GTC data from \citet{Murgas2020-WASP69b} Table A.1 (denoted {\em O1} in the following text), while the second also included the data from their Table A.2 listing the observations around the Na line (denoted as {\em O2}). Apart from the continuum information, we also wanted to put constraints on the sodium abundance, which we could then use as inputs for the high-resolution analysis with CARMENES and HARPS-N data (Sect.~\ref{sec:high-res}). To constrain the sodium volume mixing ratio, we calculated the sodium resonant-line opacities following the formalism from \citet{BurrowsEtal2000apjBDspectra}, using resonance Na~D doublet line parameters from the Vienna Atomic Line Data Base \citep{PiskunovEtal1995aapsVALDdatabase}. 

 Following the conclusion of \citet{Murgas2020-WASP69b}, we assumed that the slope seen in the WFC3/{\em HST} and OSIRIS/GTC data comes from the aerosol particles present in the planetary atmosphere and not from unocculted stellar spots. We therefore included a Rayleigh scattering model \citep{LecavelierEtal2008aaRayleighHD189733b} that allowed us to explore and constrain both the enhancement factor ($f_{\rm Ray}$) and the power-law index ($\alpha_{\rm Ray}$) of the scattering cross-section opacity. We also included a simple cloud model allowing MCMC the option to set a fully opaque cloud deck below a certain pressure (a free parameter of the model, $p_{\rm top}$).

Although several molecules have spectral features in the WFC3/{\em HST} wavelength bands in the hot-Jupiter temperature range \citep{MacDonaldMadhusudhan2017-POSEIDON}, in this analysis we considered only water as the main molecular source of opacity and retrieved its volume mixing ratio (abundance).  Our initial tests including all relevant species in the model revealed that additional species have no influence on the WASP-69~b spectra, and that the water-only model is significantly favored, with the probability ratio of $e\sp{\Delta{BIC}/2} \simeq 10^{18}$. 
This choice of near-infrared absorbers did not affect the continuum at the Na~D doublet wavelengths, which is important for our high-resolution analysis.

Water is one of the most abundant species at the WASP-69~b equilibrium temperatures explored in Table~\ref{tbl:orbital parameters} \citep[see also][]{BlecicEtal2016-TEA}. It also has the most pronounced spectral signatures at the wavelengths of our observations. To include water opacity, we used the {\tt ExoMol} molecular line-list data from \citet{PolyanskyEtal2018mnrasPOKAZATELexomolH2O}. As this database consists of billions of line transitions, we applied the {\tt Repack} package \citep{Cubillos2017apjCompress} to extract only the strongest lines that dominate the opacity spectrum between 300\,K and 3000\,K. This approach causes a difference of less than 1$\%$ in dex compared to the original line lists \citep{CubillosBlecic-PyratBayI}, but significantly improves the computational speed. Together with water and sodium opacities, our models also included collision-induced absorption from H$\sb{2}$-H$\sb{2}$ from \citet{BorysowEtal2001-H2H2highT} and \citet{Borysow2002-H2H2lowT}, and H$\sb{2}$-He from \citet{RichardEtal2012-HITRAN-CIA}.

In transmission, stellar rays travel through the atmospheric limb probing only the high altitudes (low pressures) of the planetary envelope, usually down to the 0.1\,bar level. Additionally, low-resolution observations over a narrow wavelength range such as those used in our analysis do not carry enough information to allow us to further constrain the complex temperature profiles \citep{RocchettoEtal2016apjJWSTbiases, BarstowEtal2013mnrasEchoAtmospheresCharacterization}. Thus, when retrieving the atmospheric temperature, we assumed an isothermal temperature profile.

\begin{table*}
\centering
\caption{Retrieval parameters.}
\label{tab:results}
\begin{tabular}{l l l c c}
    \hline
    \hline
    \noalign{\smallskip}
    Parameter & Symbol (unit) & Prior & \multicolumn{2}{c}{Retrieved value$^a$} \\
    \noalign{\smallskip}
    \hline
    \noalign{\smallskip}
    Planet temperature & $T_{\rm p}$ (K) & $\mathcal{U}(100,3000)$ & $900{\pm}410$ & $940{\pm}400$ \\
    Planet radius & $R_{\rm p}$ (R$_{\rm Jup}$) & $\mathcal{U}(0.8,1.2)$ & $1.01{\pm}0.02$ & $0.99{\pm}0.03$ \\
    Water abundance & $\log{{\rm H}_{2}{\rm O}}$ & $\mathcal{U}(-20,0)$ & $-2.5{\pm}1.1$ & $-2.7{\pm}1.3$ \\
    Sodium abundance & $\log{\rm Na}$ & $\mathcal{U}(-20,0)$ & ... & $-11.3{\pm}5.0$ \\
    Rayleigh strength/enhancement factor & $\log{f_{\rm Ray}}$ & $\mathcal{U}(-5,20)$ & $4.6{\pm}1.7$ & $4.3{\pm}1.6$ \\
    Rayleigh power-law index & $\log{\alpha_{\rm Ray}}$ & $\mathcal{U}(-50,0)$ & $-5.6{\pm}3.0$ & $-5.0{\pm}2.4$ \\
    Top cloud pressure & $\log{p_{\rm top}}$  (bar) & $\mathcal{U}(-8,2)$ & $-3.1{\pm}1.1$ & $-3.0{\pm}1.2$ \\
    Bayesian information criterion & BIC & ...& 63.22 & 78.55 \\
    \noalign{\smallskip}
    \hline
\end{tabular}
\tablefoot{
\tablefoottext
{a}{Mean parameter values with 1$\sigma$ uncertainties corresponding to retrievals done using the WFC3 and {\em O1} data, and WFC3 and {\em O2} data, respectively. The best-fit parameter values are given as dashed vertical lines in Figs.~\ref{fig:hist-cor} and \ref{fig:Na}, respectively.}
}
\end{table*}

In all retrievals, we also included a planet-to-star radius ratio offset parameter between the WFC3 and OSIRIS observations, $\delta_{\rm WFC3-OSIRIS}$. Offset values, on the order of those retrieved by \citet[][(see their Sect. 5]{Murgas2020-WASP69b}) have a strong and often crucial influence on the retrieved atmospheric properties. There are several possible reasons why such a $\delta_{\rm WFC3-OSIRIS}$ offset could appear between these observations: it could have originated from different system parameters adopted during the WFC3 and OSIRIS data reduction procedures \citep{AlexoudiEtal2018A, AlexoudiEtal2020}, from different instrumental systematic errors that were not fully addressed \citep[e.g.,][]{MayStevenson2020, GibsonEtal2012}, and/or from stellar activity. \citet{Murgas2020-WASP69b} explored possible values of the $\delta_{\rm WFC3-OSIRIS}$ offset due to the different system parameters, and calculated the remaining offset assuming that it comes from stellar activity. Our analysis in Sect. \ref{sec:long-term}, Fig. \ref{fig:photometry_stella}, and Sect. \ref{sec:stellar_activity} also confirms activity in WASP-69 star.

To address this, we initially allowed $\delta_{\rm WFC3-OSIRIS}$ to be a free parameter in our model. However, none of our retrievals was able to put constraints on this parameter, at the same time causing multiple correlations between other free parameters (we explore these correlations and degeneracies in our following paper \citealt{BlecicEtal2021-WASP-69b}). Here, we adopted $\delta_{\rm WFC3-OSIRIS} = 479$\,ppm (we shift the WFC3 data upwards), which corresponds to the offset retrieved by \citet[][see their Table~4]{Murgas2020-WASP69b} for their case without stellar spots. Our decision to choose this value is based on the fact that it is comparable to the variability of the stellar flux of $\sim$2.4\,\% found by \citet{Anderson2014}, assuming that such variability occurs in regions not occulted by the planet. We note that by adopting a fixed offset in this analysis, we neglected some uncertainties caused by correlations and degeneracies.

Our model atmosphere had 50 equally spaced layers in log pressure, ranging from 10$^{2}$\,bar to 10$^{-8}$\,bar, and we included all molecular species relevant for the hydrogen-dominated atmospheres with solar C/O ratio (H$_2$, He, H$_2$O, CH$_4$, CO, CO$_2$, HCN, N$_2$, NH$_3$), as well as sodium. We used the {\tt TEA} code \citep{BlecicEtal2016-TEA} to calculate the chemical equilibrium abundances of all the species at 1000\,K and used them for the initial model atmosphere assuming that species abundances remain constant with altitude. In retrieval, we leave H$_2$O abundance as the only free parameter in our model, while we kept other species abundances at initial values. We set the transit planetary radius at 0.01\,bar pressure level and calculated the altitude profile assuming hydrostatic equilibrium. The system parameters used to generate the spectra are listed in Table~\ref{tbl:orbital parameters}. 

We constrained the atmospheric parameters listed in Table~\ref{tab:results} assuming wide uniform or log-uniform priors for all parameters. We explored the posterior parameter space with the {\tt Snooker} differential-evolution MCMC code  \citep[DEMC,][]{terBraak2006DifferentialEvolution, terBraak2008SnookerDEMC}, and obtained between two and six million samples (after discarding the initial 10\,000 iterations). This ensured that the \citet{GelmanRubin1992} statistics remained at $\sim$1.01 or lower for each free parameter. To determine the best-fit model among those generated on the same dataset, we used the Bayesian information criterion, BIC = $\chi^2 + k\,\ln{N}$.


\begin{figure*}[t!]
    \centering
    \includegraphics[width=0.65\textwidth]{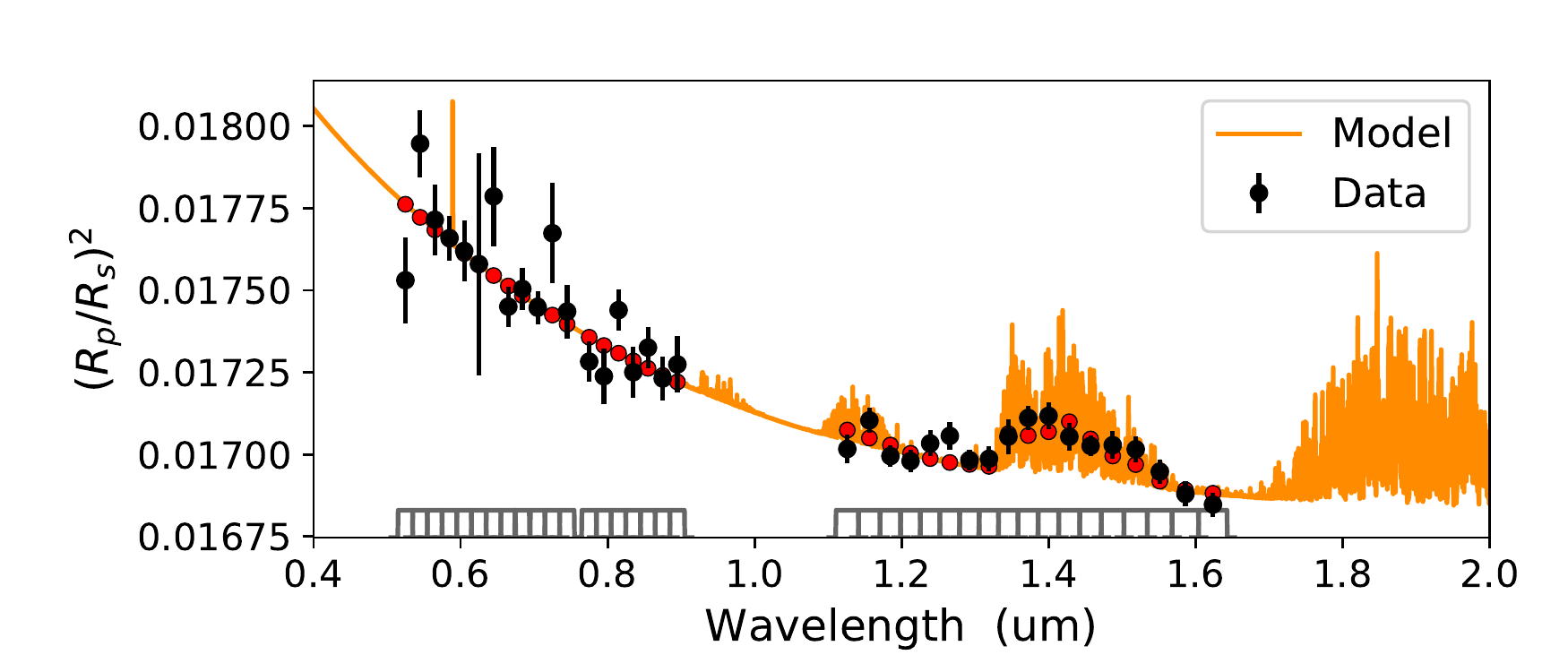}\hspace{-10pt}
    \caption{\textit{Left:} Best-fit spectrum when the WFC3 and {\em O1} are included. Black points denote the data with uncertainties, while the red points denote the best-fit spectrum integrated over the bandpasses of the observations, which are shown in gray at the bottom of the panel. }
    \label{fig:spectrum}
\end{figure*}


\begin{figure*}[t!]
    \centering
    \includegraphics[width=0.40\textwidth]{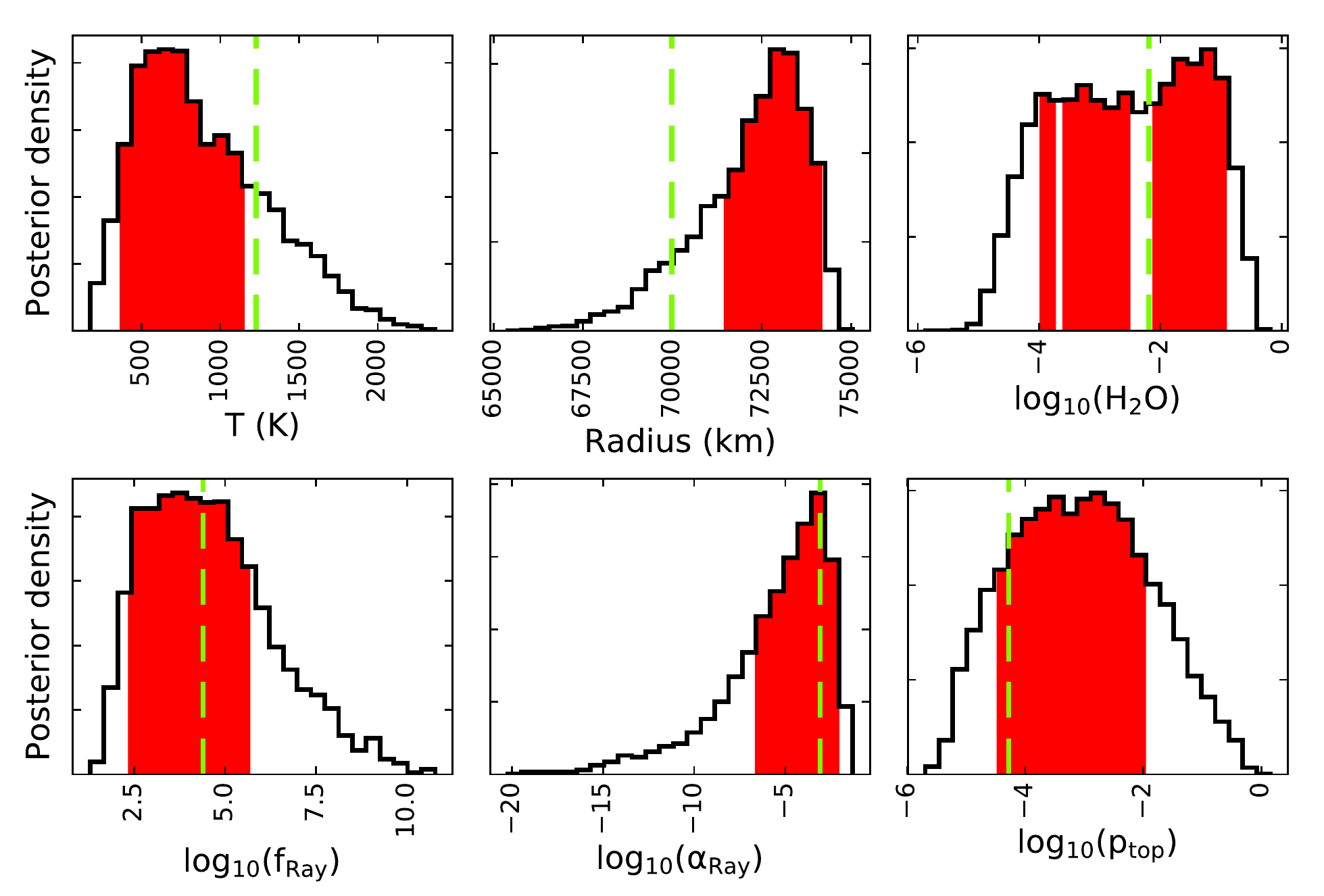}
    \includegraphics[width=0.50\textwidth]{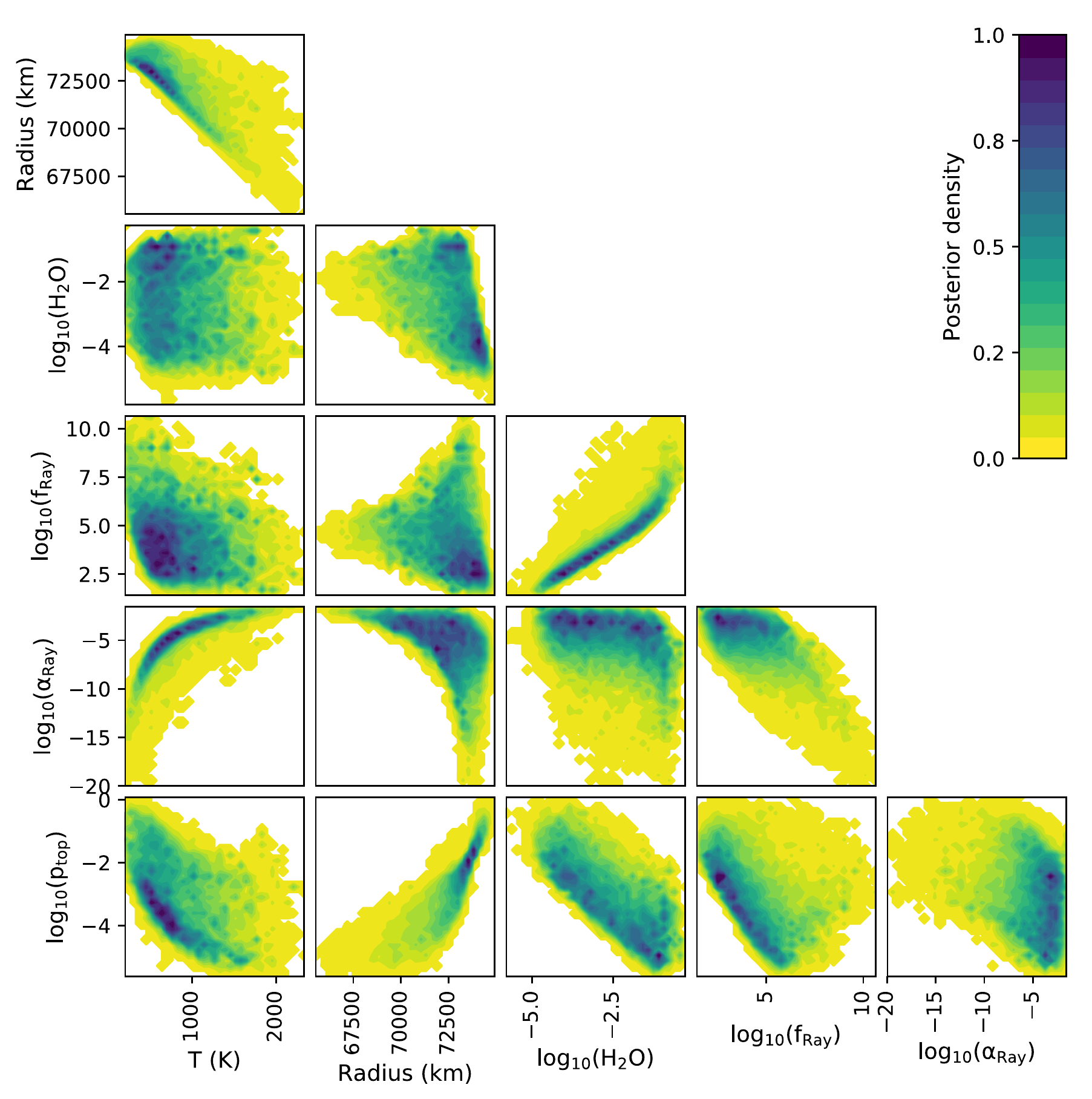}
    \caption{\textit{Left:} Marginal posterior distributions (histograms) when the WFC3 and {\em O1} data are included. The shaded red areas denote the 68\,\% confidence intervals of the respective distributions. The light green dashed vertical lines denote the retrieved best-fit parameter values. Strong non-linear correlations seen in the right panel of this figure produce that some of the best-fit parameter values fall outside of the red credible regions, and cause gaps in the water confidence interval, reflecting a bi-modal posterior distribution.
\textit{Right:} Pair-wise correlation plots with posterior densities.}
    \label{fig:hist-cor}
\end{figure*}

\subsubsection{Retrieval results}

\begin{figure*}
    \centering
    \includegraphics[width=0.49\textwidth]{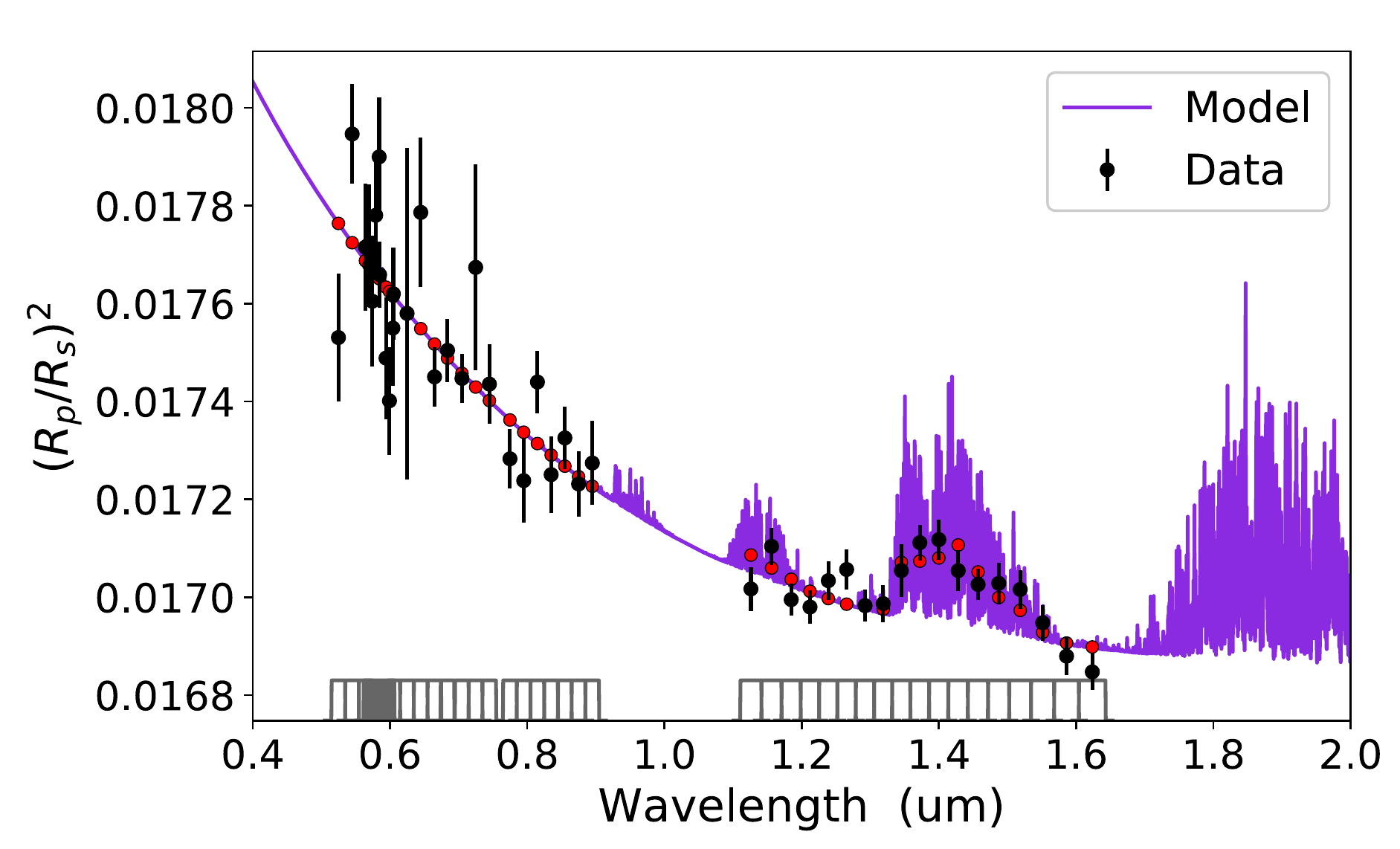}
    \includegraphics[width=0.49\textwidth]{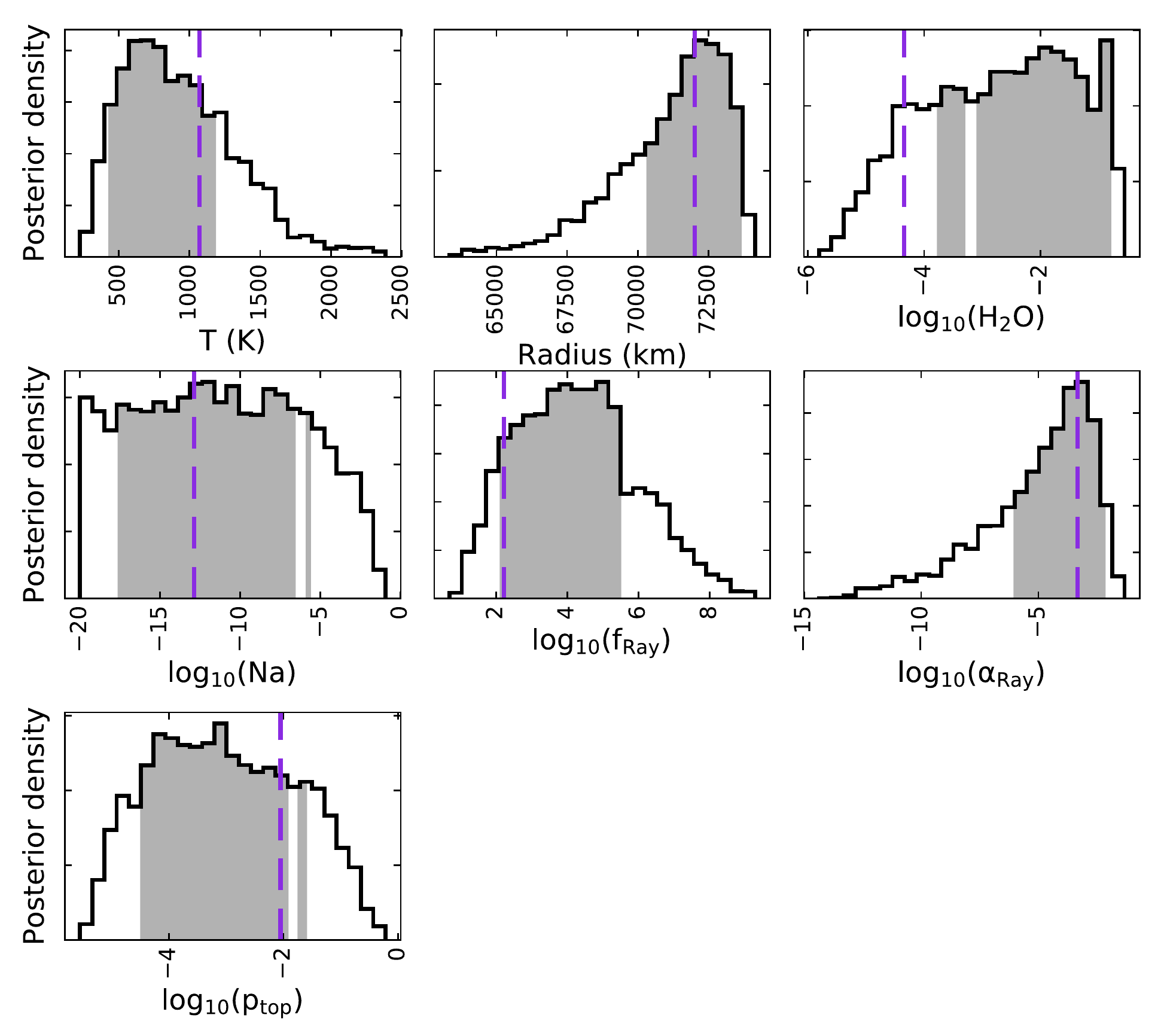}
    \caption{\textit{Left:} Best-fit spectrum when the WFC3 and {\em O2} data are included. Black points denote the data with uncertainties, while the red points denote the best-fit spectrum integrated over the bandpasses of the observations, which are shown in gray at the bottom of the panel.  \textit{Right:} Marginal posterior distributions (histograms). The shaded gray areas denote the 68\,\% confidence intervals of the respective distributions. The purple dashed vertical lines denote the retrieved best-fit parameter values. The pair-wise correlation plots are similar to those shown in Figure~\ref{fig:hist-cor}, displaying same nonlinear correlations and degeneracies.}
    \label{fig:Na}
\end{figure*}

Figure~\ref{fig:spectrum} shows the best-fit spectrum model for the case of WFC3 and {\em O1} data. In this case, we retrieved the planet temperature, radius, water abundance, Rayleigh slope, and strength factor, and the top pressure of an opaque cloud deck. The mean parameter values with 1$\sigma$ uncertainties are given in Table~\ref{tab:results}, while the best-fit parameter values are shown in the posterior density plots in the left panel  of Fig.~\ref{fig:hist-cor}. In the right panel of Fig.~\ref{fig:hist-cor}, we show the pair-wise correlation plots with posterior densities. Figure~\ref{fig:Na} shows the best-fit model and posterior density plots for the case of WFC3 and {\em O2} data. In this case, in addition to the parameters listed above, we also retrieved the sodium abundance. Figures~\ref{fig:spectrum} and~\ref{fig:Na} show good fits to the data for both cases.

We note a steep Rayleigh slope from $\sim1.2$\,$\mu$m towards shorter wavelengths. The transmission functions confirm that the atmosphere becomes opaque above 10$^{-4}$\,bar for the near-infrared observations, and above 10$^{-5}$\,bar for the optical observations. This behaviour is expected, as Rayleigh scattering becomes very pronounced below 1\,$\mu$m when aerosol particles with sizes smaller than or equal to the wavelength of light cause the atmosphere to become opaque for the incoming radiation. Rayleigh opacity usually scales with $\lambda \sp{-4}$. \citet{MallonnWakeford2017} performed a comparative study of the scattering slopes of all planets in \citet{Sing2016} together with \object{HAT-P-32}~b and \object{GJ~3470}~b, and found that the majority of them have slope values $\log{\alpha_{\rm Ray}} > -4$ (between --1 and --3), with a couple of outliers (\object{HD~189733}~b and \object{GJ~3470}~b) having $\log{\alpha_{\rm Ray}} < -5$. Later studies by \citet{OhnoKawashima2020-Rayleigh} and \citet{MayEtal2020} found that Rayleigh slopes can indeed be steeper than --4. These latter authors discussed the idea that opacity gradients can be naturally generated in hot Jupiters with species other than H, H$\sb{2}$, and He, such as photochemical hazes, when eddy mixing in the planetary atmosphere is very efficient. These findings are in line with those of \citet{PinhasEtal2019} and \citet{WelbanksEtal2019}, who obtained a median spectral index of $\log{\alpha_{\rm Ray}} \lesssim -5$ for different sets of hot Jupiter planets. Both of our analyses (Table~\ref{tab:results} and Figs.~\ref{fig:hist-cor} and~\ref{fig:Na}) reveal a Rayleigh slope on the order of $\log{\alpha_{\rm Ray}} \sim -5$, steeper than the slope caused by the tiny aerosol particles.


None of our retrievals returned strong constraints on the planetary temperature. The  left panel of Figure \ref{fig:hist-cor} shows that the 1$\sigma$ region spans around 800 K (red section, top left panel). The mean value, however, is close to the planetary equilibrium temperature, assuming albedo $A_{\rm Bond}=0$ and efficient energy redistribution. The radius of the planet and the Rayleigh slope are most robustly constrained, revealing a planet with a radius very close to that of Jupiter and strong Rayleigh scattering.

The water abundance seems to be slightly bi-modal (solar and super-solar) in the first case (Fig.~\ref{fig:hist-cor}, {\em O1} data), with a tendency towards large values, $\log{{\rm H}_{2}{\rm O}} \sim -2$, and loosely constrained in the second case (Fig.~\ref{fig:Na}, {\em O2} data), spreading over several orders of magnitude between 10$\sp{-4}$ (solar) to 10$\sp{-1}$ (super-solar) in volume mixing ratio. WASP-69~b has a Jupiter-like radius, but a very small mass (0.26\,M$_{\rm Jup}$). 
\citet{WelbanksEtal2019} showed that there is a metallicity trend of increasing water abundances with decreasing mass going from gas-giant planets to mini-Neptunes. In particular, low-mass planets such as HATP-26~b, WASP-39~b, and WASP-127b show super-solar water abundance on the order of $\log{{\rm H}_{2}{\rm O}} \sim -2$ and smaller. \citet{WelbanksEtal2019} suggested that this trend could be a consequence of the different formation pathways from those usually assumed for hot Jupiter planets.

In the second case ({\em O2} data), we also added sodium as an additional source of opacity and allowed its abundance to be a free parameter of the model. This retrieval returned results akin to those in the first case, with most of the parameters having similar values and confidence ranges (Figs.~\ref{fig:spectrum} and~\ref{fig:Na}; see also Table \ref{tab:results}). The spectrum, temperature posteriors, transmission functions, and correlation plots for the second case also looked comparable to the plots from the first case (Figs.~\ref{fig:spectrum} and~\ref{fig:hist-cor}). The sodium abundance, of interest here, is loosely constrained to $\log{\rm Na} \simeq -11$. Considering that this abundance is negligible, causing very low sodium opacity, the best-fit model in Fig.~\ref{fig:Na} does not show any trace of the sodium doublet in the spectrum. Sodium, if present and abundant, would otherwise have a distinct signature at 5895\,\AA.

Our results (Table \ref{tab:results}) agree well with results from previous analyses (Section \ref{sec:model}), and in particular with the results of \citet{Murgas2020-WASP69b} who used the same datasets. Our retrieved planetary radius, temperature, and Rayleigh parameters are consistent with theirs. \citet{Murgas2020-WASP69b} also found a high atmospheric metallicity (together with a low C/O ratio compared to solar), causing a super-solar water abundance on the order of our retrieved value.
 
\subsection{High-resolution atmospheric modeling}
\label{sec:high-res}

\begin{figure*}
    \centering
    \includegraphics[width=0.99\textwidth]{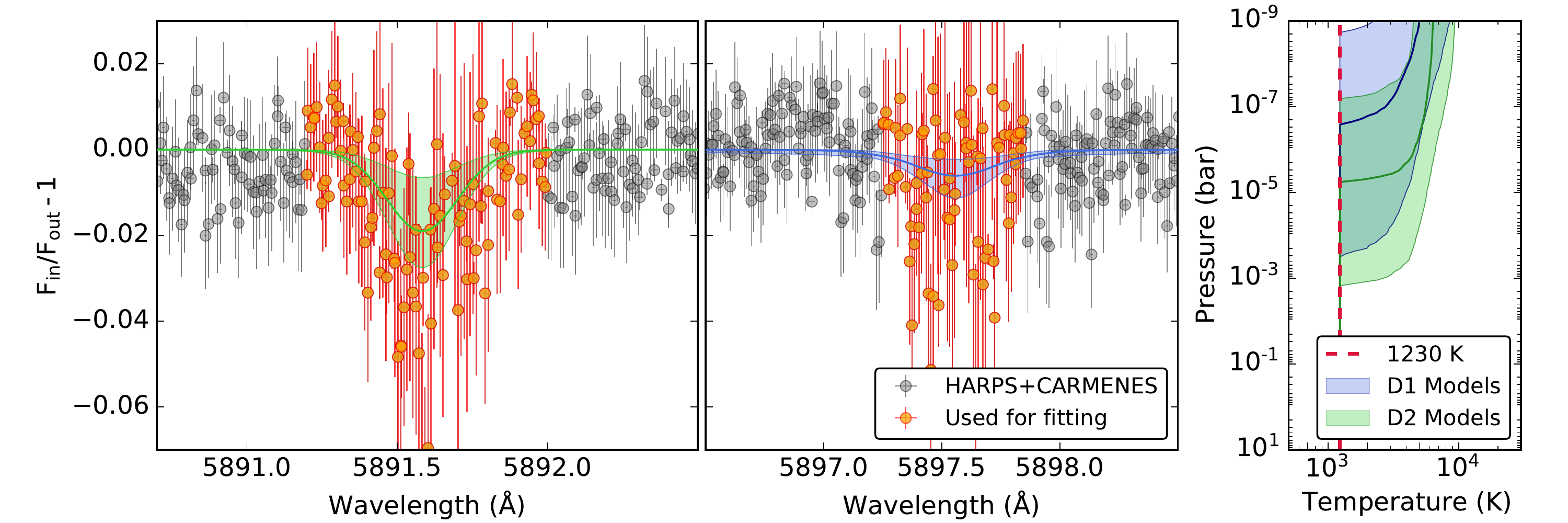} 
    \includegraphics[width=0.75\textwidth]{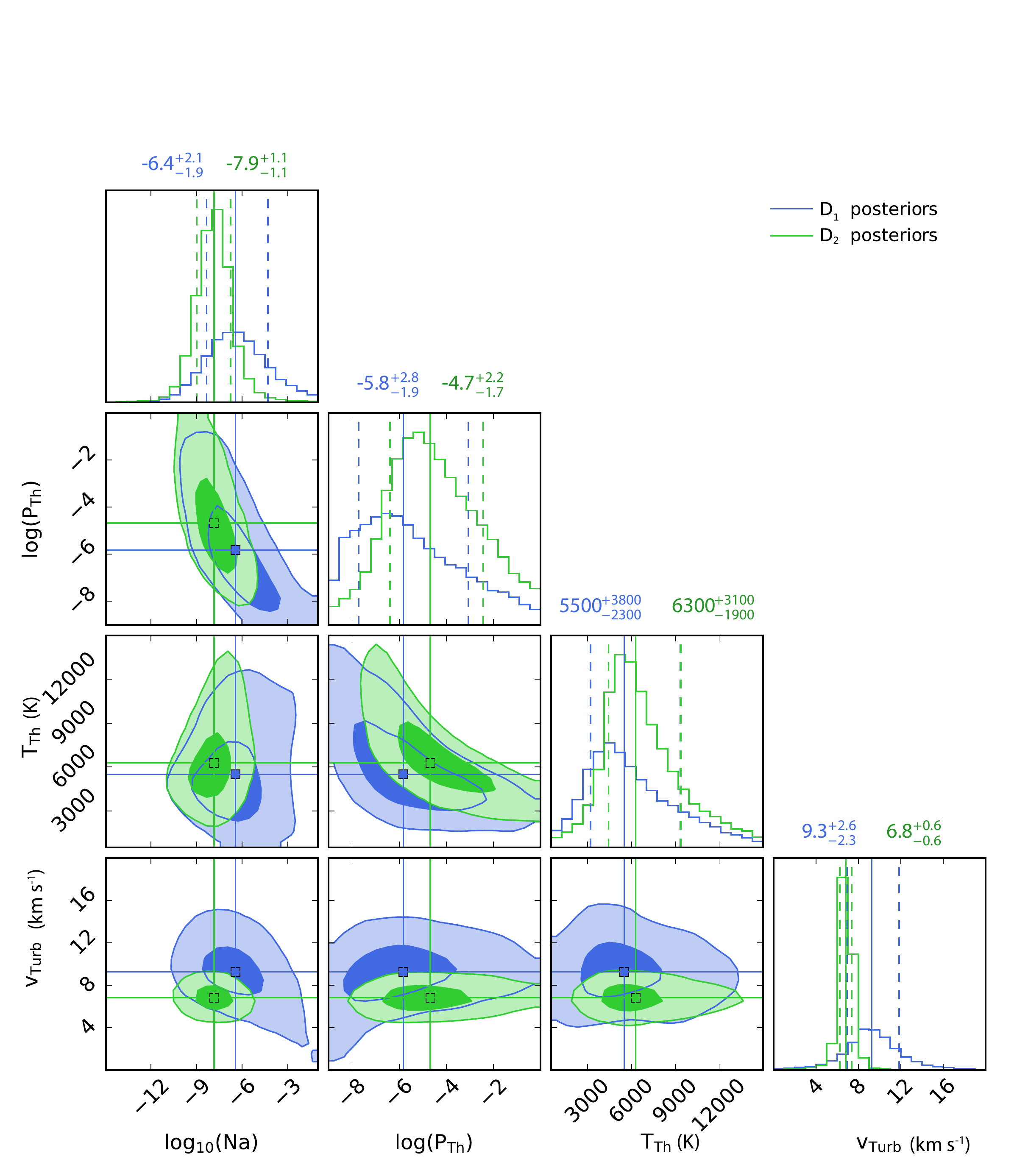}
    \caption{{\em Top}: Bayesian inference of the upper thermospheric conditions of WASP-69~b. 
    Comparison of the best-fit model spectra of Na~{\sc i} D$_{2}$ ({\em left panel}, shaded green area) and Na~{\sc i} D$_{1}$ ({\em middle panel}, shaded blue area) with discarded and used high-resolution spectroscopic data in gray and orange, respectively, and retrieved temperature profiles based on the posteriors ({\em right panel}). 
    The dashed vertical line in red marks the retrieved temperature using low-resolution spectra (Sect.~\ref{sec:low-res}). 
    Uncertainties are marked by vertical dashed lines at the 16\,\%, 50\,\%, and 84\,\% quantiles. {\em Bottom}: Posterior probability distribution of high-resolution model parameters as well as the marginalized distribution for each parameter for Na~D$_{2}$ (shaded green area) and D$_{1}$ (shaded blue area) in WASP-69~b. The marginalized uncertainties are marked by vertical dashed lines at the 16\,\%, 50\,\%, and 8\,4\% quantiles. In the density maps, $1\sigma$ and $2\sigma$ are given as $1-e^{-0.5}\sim 0.393$ and $1-e^{-2.0}\sim 0.865$, respectively, as is common for multivariate MCMC results.} 
    \label{fig:high-res-model}
\end{figure*}

Our analysis in Sect.~\ref{sec:discuss_Na} and \ref{sec:discuss_stat_Na} suggested the feasibility of teasing out information from the current high-resolution datasets. 
Likewise, in Sect.~\ref{sec:low-res} we show that the information content in the low-resolution spectrum is statistically significant. 
In this section, we discuss steps in order to characterize the upper thermosphere of WASP-69~b using atmospheric parameters retrieved from low-resolution datasets as input, and combining them with the CARMENES (first night) and HARPS-N observations.

Normalization of the continuum is a common step in the data reduction of high-resolution spectra of exoplanetary atmospheres. 
However, such normalization comes at the price of losing continuum information, which in turn causes additional degeneracies in the retrieved values \citep{fisher2019much}. 
A common degeneracy in high-resolution studies occurs between the retrieved abundance of the species, reference pressure, and atmospheric temperature \citep[e.g.][]{brogi2019retrieving}, where different combinations of these parameters could result in similar line profiles. This issue becomes particularly significant when the S/N is relatively low. However, a precise line shape at high S/N could break the degeneracies to some extent \citep[e.g.,][; see also \citet{welbanks2019mass} for a similar discussion on low-resolution retrieval degeneracies]{benneke2012atmospheric,heng2017theory,brogi2019retrieving,gibson2020detection}. 
To break these degeneracies and estimate the continuum at the Na~D doublet, we employed the model parameters from low-resolution spectra as estimated in the previous section (see the parameters best-fit values in histograms of Fig. \ref{fig:Na} and \ref{fig:hist-cor}) as priors for the high-resolution atmospheric model.

We used a Python implementation of {\tt emcee} \citep{goodman2010ensemble,foreman2013emcee}, called {\tt MCKM}, to effectively sample the parameter-space. 
{\tt MCKM} handles any arbitrary number of model parameters over regular (Cartesian) or irregularly spaced steps and any arbitrary number of data points and their respective covariance matrices. 
The observational uncertainties can be estimated through a Gaussian process using a variety of kernels. 
We assumed uninformative priors to initialize 1000 walkers in the MCMC process and iterated until the solution converged.

Table~\ref{tbl:prior+posterior} summarizes the model parameters, that is, Na abundance, thermospheric onset pressure, thermosphere temperature, effective turbulent velocity, and their priors. 
We used the core radiative transfer solver of {\tt petitRADTRANS}  \citep{molliere2019petitradtrans} to calculate forward radiative transfer models and included broadening due to turbulent velocity. In order to model the effect of turbulent velocity, we assume a Gaussian velocity distribution, assuming the effective turbulent velocity as the variance of the distribution and zero bulk velocity. We set up the high-resolution model by discretizing the 1D atmosphere to 150 layers, linearly spaced in altitude. We set the lower boundary at 1\,bar and the upper boundary at $10^{-9}$\,bar to resolve the centroids in grazing geometry appropriate for transmission spectroscopy.

We also clipped the high-resolution data to exclude data points far away from the signal and speed up the retrieval. 
The clipped data are shown in Fig.~\ref{fig:high-res-model}. 
Including the discarded points did not change the results and the posteriors were robust against the choice of clipping wavelengths outside the used area.

To explain large D$_{2}$/D$_{1}$ ratios (i.e., 2.5$\pm$0.7; see Section~\ref{sec:discuss_Na}), it has been hypothesized that D$_{2}$ and D$_{1}$ may probe different thermosphere regions with different temperatures. 
Other possibilities are the effect of nonlocal thermodynamic equilibrium (nonLTE) processes \citep{lind2011non} or an optically thin sodium contribution in the composition of the thermosphere \citep{Gebek2020}. 
We examined these hypotheses by fitting our radiative transfer model to the D$_{2}$ and D$_{1}$ lines separately and by comparing the posterior distributions of the resulting model parameters.

Table~\ref{tbl:prior+posterior} also summarizes the posterior distributions of the parameters and Fig.~\ref{fig:high-res-model} illustrates the best-fitted models to the data as well as our retrieved posterior probability distributions of the model parameters.
We find that the retrieved parameter values are statistically consistent, which suggested that both lines may originate from the same region and at similar temperatures. 
Also, given our results, there was no need for nonLTE processes to explain the observed D$_{2}$/D$_{1}$ ratio.

Our retrieved sodium abundances from the D$_{2}$ and D$_{1}$ lines suggest that the majority of their absorption occurs at optically thin regimes. 
This scenario is consistent with the conclusion of \citet{Gebek2020} because our finding supports their main assumption (i.e., both D$_{2}$ and D$_{1}$ lines probe the same region and arise from the same sodium population) and also supports the optically thin condition for the sodium lines (i.e., our retrieved low abundances). 
However, it is unclear if the conclusion of these latter authors, and hence our comparison, would hold true if D$_{2}$ and D$_{1}$ probed different regions of the atmosphere. 
Moreover, \citet{baranovsky2004photon} showed that the D$_{2}$/D$_{1}$ ratio could be significantly affected by the multiple scattering under optically thick conditions and the D$_{2}$/D$_{1}$ ratio could decrease to very small values. 
Further observations are required to reach a higher S/N and to further constrain the posteriors and address this question.

The posteriors on the thermosphere onset pressure suggest that both D$_{2}$ and D$_{1}$ lines probe pressures lower than the millibar regime.
Our retrieved temperatures also appear to be consistent with 6000\,K as inferred for the thermosphere of the ultra-hot Jupiter \object{WASP-189}~b \citep[e.g.][]{yan2020temperature}.  
However, WASP-69~b is not as highly irradiated as WASP-189~b and their thermospheres might differ in nature. More data are needed to reduce the uncertainties in thermosphere temperatures to ascertain more reliable thermosphere properties. It is worth noting that obtaining constraints on the thermosphere of exoplanets via high-resolution spectroscopy is a topic at the forefront of exoplanet science and perhaps obtaining some constraints on the upper atmosphere for such a relatively cool exoplanet could set the stage for subsequent investigation and the improvement in characterization of exoplanet thermospheres.

We also find that an effective turbulent velocity on the order of 5--10\,km\,s$^{-1}$ is necessary to explain the line broadening in both the D$_{2}$ and D$_{1}$ lines. 
These values are in line with the velocities reported by \citet{Lampon_modellng_2021}, which suggests consistency in retrieved upper thermospheric velocities on irradiated gaseous exoplanets. 
However, these velocities are slightly higher than the velocities estimated by 3D general circulation models \citep[e.g.,][]{showman2010atmospheric,pierrehumbert2019atmospheric,wang2020extremely}, which might hint at the existence of other broadening mechanisms in addition to the rotation broadening. 
Nevertheless, all these values are considerably lower than the 40\,km\,s$^{-1}$ reported by \citet{Seidel2020}. 
Again, higher S/N spectra are required to unambiguously investigate the nature of these broadenings.

\begin{table*}
\centering
\caption {\label{tbl:prior+posterior} Prior and posterior parameters of the high-resolution model fit.}
\begin{tabular}{l l l c c}
\hline\hline
\noalign{\smallskip}
Parameter & Symbol (unit) & Prior & \multicolumn{2}{c}{Posterior} \\
~ & ~ & ~ & Na~{\sc i} D$_1$ & Na~{\sc i} D$_2$ \\
\noalign{\smallskip}
\hline
\noalign{\smallskip}
Sodium solar fraction & $\log{\rm Na}$  & $\mathcal{U}$(--15,--1)    & $-6.4^{+2.1}_{-1.9}$    & $-7.9^{+1.1}_{-1.1}$ \\
\noalign{\smallskip}
Thermospheric onset pressure & $\log{P_{\rm th}}$ (bar)  & $\mathcal{U}$(--9,0)         & $-5.8^{+2.8}_{-1.9}$ & $-4.7^{+2.2}_{-1.7}$ \\
\noalign{\smallskip}
Thermospheric temperature & $T_{\rm th}$ (K)     & $\mathcal{U}$(500,15000)      & $5500^{+3800}_{-2300}$  &  $6300^{+3100}_{-1900}$ \\
\noalign{\smallskip}
Effective turbulent velocity & $v_{\rm turb}$ (km\,s$^{-1}$) & $\mathcal{U}$(0,20)  & $9.3^{+2.6}_{-2.3}$ & $6.8^{+0.6}_{-0.6}$ \\ 
\noalign{\smallskip}
\hline
\end{tabular}
\tablefoot{
\tablefoottext
{a}{
In the prior parameters, $\mathcal{U}$($a$,$b$) stands for a uniform distribution between $a$ and $b$.
The uniform distributions in $\log{\rm Na}$ and $\log{P_{\rm th}}$ are equivalent to Jeffrey’s log-uniform distributions of Na and $P_{\rm th}$ between $10^{a}$ and $10^{b}$ (i.e., $\mathcal{J}$($10^{-15},10^{-1}$) and $\mathcal{J}$($10^{-9},10^{0}$), respectively).
The posterior parameters are based on 16\,\%, 50\,\%, and 84\,\% quantiles.}
}
\end{table*}

\section{Summary}
\label{sec:conclusions}

In this work, we performed a combined high- with low-resolution transmission spectroscopy analysis to obtain a more complete picture of the properties of the atmosphere of WASP-69 b. Particularly, we aimed to break the degeneracy in abundance and atmospheric temperature, which is present in any ground-based high-resolution-only modeling due to lack of information on the continuum of the transmission spectrum.
We presented CARMENES optical and near-infrared spectroscopy during three transits of the low-density hot Saturn WASP-69~b.
We searched for planetary atmospheric features from 0.52\,$\mu$m to 1.71\,$\mu$m using the transmission spectroscopy technique, with special emphasis on the Na~{\sc i} D$_1$ and D$_2$ doublet.
We also added to our analysis the available low-resolution data of this target, as well as the transmission spectra obtained with HARPS-N, and performed atmospheric retrieval on both low- and high-resolution data. 
We used the fitting parameters of our low-resolution model as priors for our high-resolution modeling. 

Due to insufficient data quality and a high level of contamination by sky emission, we disregarded the transit data of the second and third nights in our analysis of the Na D lines.
Using the first night data only, we measured Gaussian amplitudes of 3.2\,$\pm$\,0.3\,\% (D$_{2}$) and 1.2\,$\pm$\,0.3\,\% (D$_{1}$), and $\sigma$ of 0.13\,$\pm$\,0.01\,{\AA} (D$_{2}$) and 0.14\,$\pm$\,0.2\,{\AA} (D$_{1}$).
These values correspond to detection at 7$\sigma$ and 3$\sigma$ significance at Na~{\sc i} D$_{2}$ and D$_{1}$, respectively. Our sodium detection is in agreement with previous measurements \citep{Casasayas-Barris2017}.
In addition, although strong signatures of an escaping atmosphere were observed from the upper atmosphere of WASP-69~b through investigation of the He~{\sc i} triplet lines \citep{Nortmann2018}, we found no clear signature of escape in the optical region of the transmission spectrum.

In the analysis of the H${\alpha}$ line, only the third night data showed some hints of this feature. However, further investigations with data of better quality are required to confirm this feature.

Our low-resolution atmospheric analysis revealed a Jupiter-radius planet with strong Rayleigh scattering, solar to super-solar water abundance, and a highly muted sodium feature, in agreement with the findings of previous low-resolution studies by \citet{Murgas2020-WASP69b}.
The parameters of this model related to the continuum were used in our high-resolution modeling.

In our high-resolution observations, we measured a D$_{2}$/D$_{1}$ ratio of 2.5\,$\pm$\,0.7 and investigated the possible physical conditions that lead to this result.
Further, in our high-resolution atmospheric modeling, we fit  the D$_{2}$ and D$_{1}$ lines individually and compared their atmospheric parameter posteriors. 
We find that the retrieved temperatures ($T_{\rm th}\sim$ 6000$\pm$3000\,K) and abundances ($\log{\rm Na}\sim 7\pm2$) of these two lines agree with each other to better than 1$\sigma$. 
Therefore, the observed high D$_{2}$/D$_{1}$ ratio is consistent with LTE conditions and the non- or weak detection of D$_{1}$ is a consequence of obscuration of this line by optical opacity (i.e., haze). We also find that an effective turbulent velocity of 5--10\,km\,s$^{-1}$ is needed to explain the line broadening in both D$_{2}$ and D$_{1}$.

\begin{acknowledgements}
We thank G.~Zhou, J.~Seidel and A. Wyttenbach for a fruitful scientific discussion. 

CARMENES is an instrument at the Centro Astron\'omico Hispano-Alem\'an (CAHA) at Calar Alto (Almer\'{\i}a, Spain), operated jointly by the Junta de Andaluc\'ia and the Instituto de Astrof\'isica de Andaluc\'ia (CSIC).
  
CARMENES was funded by the Max-Planck-Gesellschaft (MPG), 
the Consejo Superior de Investigaciones Cient\'{\i}ficas (CSIC),
the Ministerio de Econom\'ia y Competitividad (MINECO) and the European Regional Development Fund (ERDF) through projects FICTS-2011-02, ICTS-2017-07-CAHA-4, and CAHA16-CE-3978, 
 and the members of the CARMENES Consortium 
  (Max-Planck-Institut f\"ur Astronomie,
  Instituto de Astrof\'{\i}sica de Andaluc\'{\i}a,
  Landessternwarte K\"onigstuhl,
  Institut de Ci\`encies de l'Espai,
  Institut f\"ur Astrophysik G\"ottingen,
  Universidad Complutense de Madrid,
  Th\"uringer Landessternwarte Tautenburg,
  Instituto de Astrof\'{\i}sica de Canarias,
  Hamburger Sternwarte,
  Centro de Astrobiolog\'{\i}a and
  Centro Astron\'omico Hispano-Alem\'an), 
with additional contributions by the MINECO, 
the Deutsche Forschungsgemeinschaft (DFG) through the Major Research Instrumentation Programme and Research Unit FOR2544 ``Blue Planets around Red Stars'', 
the Klaus Tschira Stiftung, 
the states of Baden-W\"urttemberg and Niedersachsen, 
and by the Junta de Andaluc\'{\i}a.
  
We acknowledge financial support from the DFG through priority program SPP 1992 ``Exploring the Diversity of Extrasolar Planets'' (KH 472/3-1) and through grant CA 1795/3,
NASA through ROSES-2016/Exoplanets Research Program (NNX17AC03G),
the Klaus Tschira Stiftung,
the European Research Council under the European Union's Horizon 2020 research and innovation program (694513),
the Agencia Estatal de Investigaci\'on of the Ministerio de Ciencia, Innovaci\'on y Universidades and the ERDF through projects 
  PID2019-109522GB-C5[1:4]  
and the Centre of Excellence ``Severo Ochoa'' and ``Mar\'ia de Maeztu'' awards to the Instituto de Astrof\'isica de Canarias (SEV-2015-0548), Instituto de Astrof\'isica de Andaluc\'ia (SEV-2017-0709), and Centro de Astrobiolog\'ia (MDM-2017-0737), and the Generalitat de Catalunya/CERCA programme.

\end{acknowledgements}


\bibliography{bib}

\newpage
\appendix

\section{Long-term photometric observation}

\begin{figure*}
    \centering
    \includegraphics[scale = 0.5]{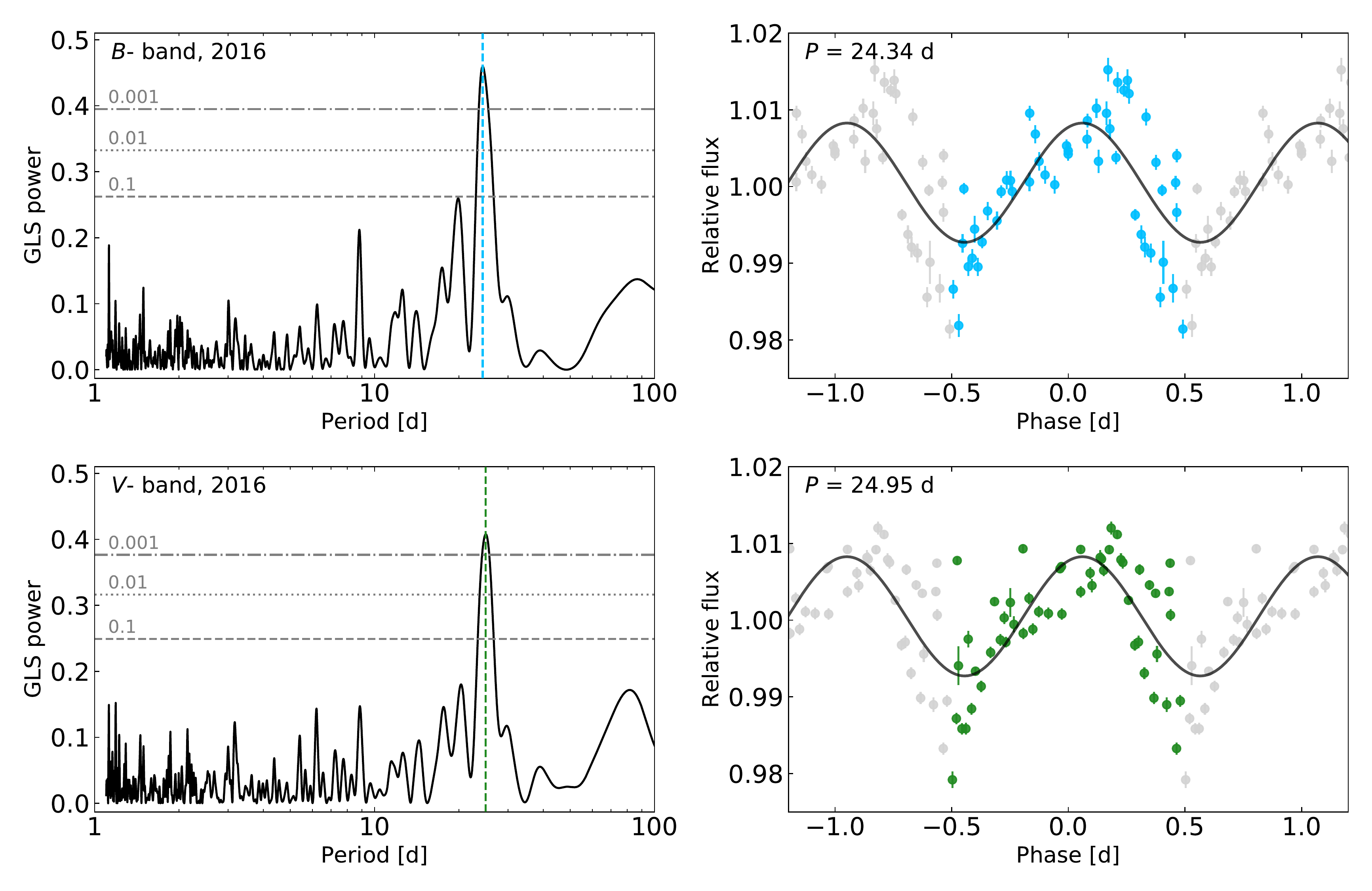}
    \caption{{\em Left panels:} GLS periodograms for the WASP-69 data in $B$ and $V$ bands from the 2016 observing season. Horizontal lines indicate false-alarm probability levels FAP = 0.1, 0.01, and 0.001. {\em Right panels:} Light curves folded to the best periods, as noted in the plots. A sine wave for the best periods is fitted to the data. The blue and green colors show one complete cycle in $B$ and $V$ filters, respectively.}
    \label{fig:per-gram}
\end{figure*}

Periodograms and phase-folded light curves related to the photometric observations in 2016 in $B$ and $V$ bands are shown in Fig.~\ref{fig:per-gram}

\section{Analysis of the data from the second and third nights}
\label{sec:appen1}

\begin{figure}
    \centering
    \includegraphics[width=0.95\columnwidth]{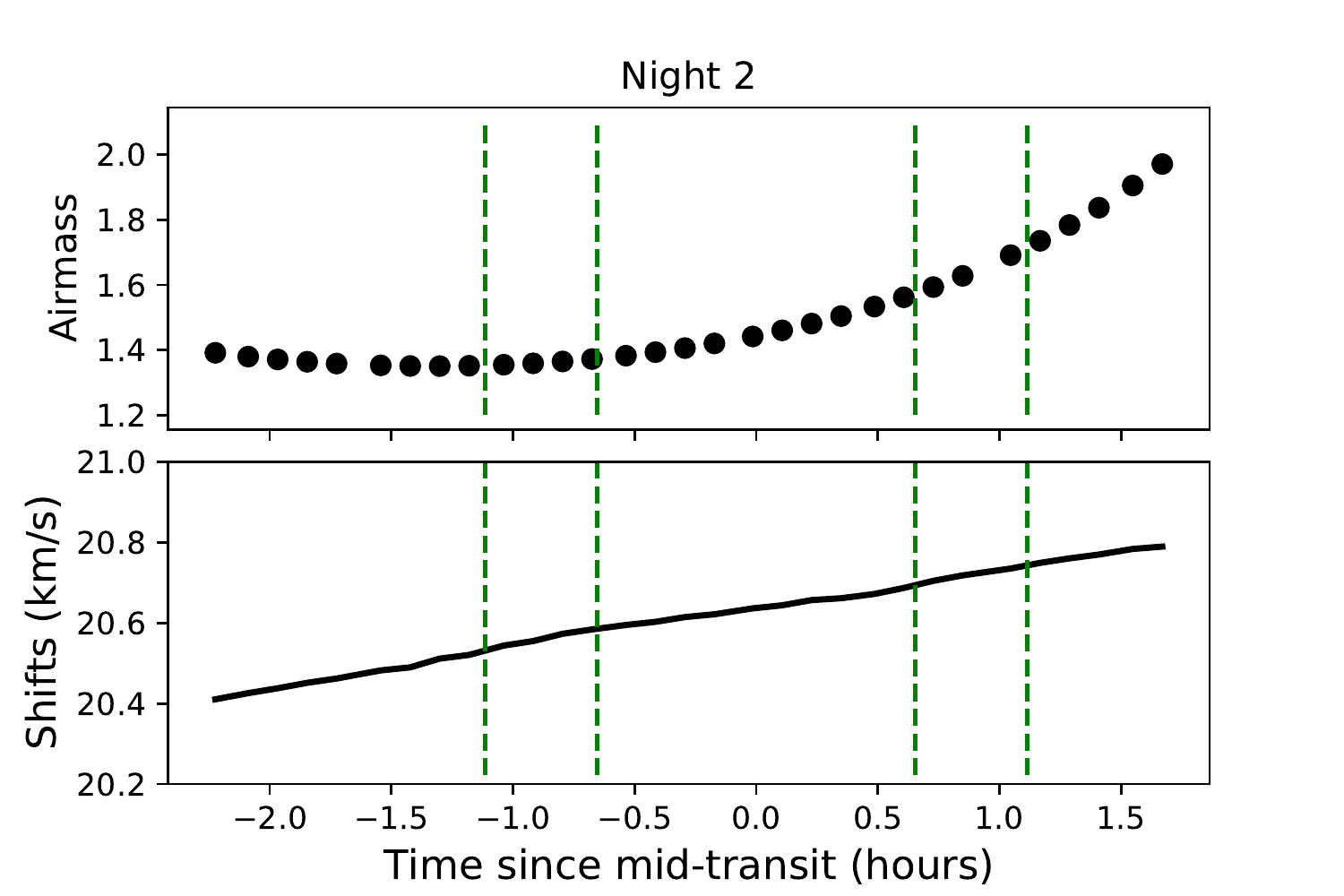}
    \includegraphics[width=0.95\columnwidth]{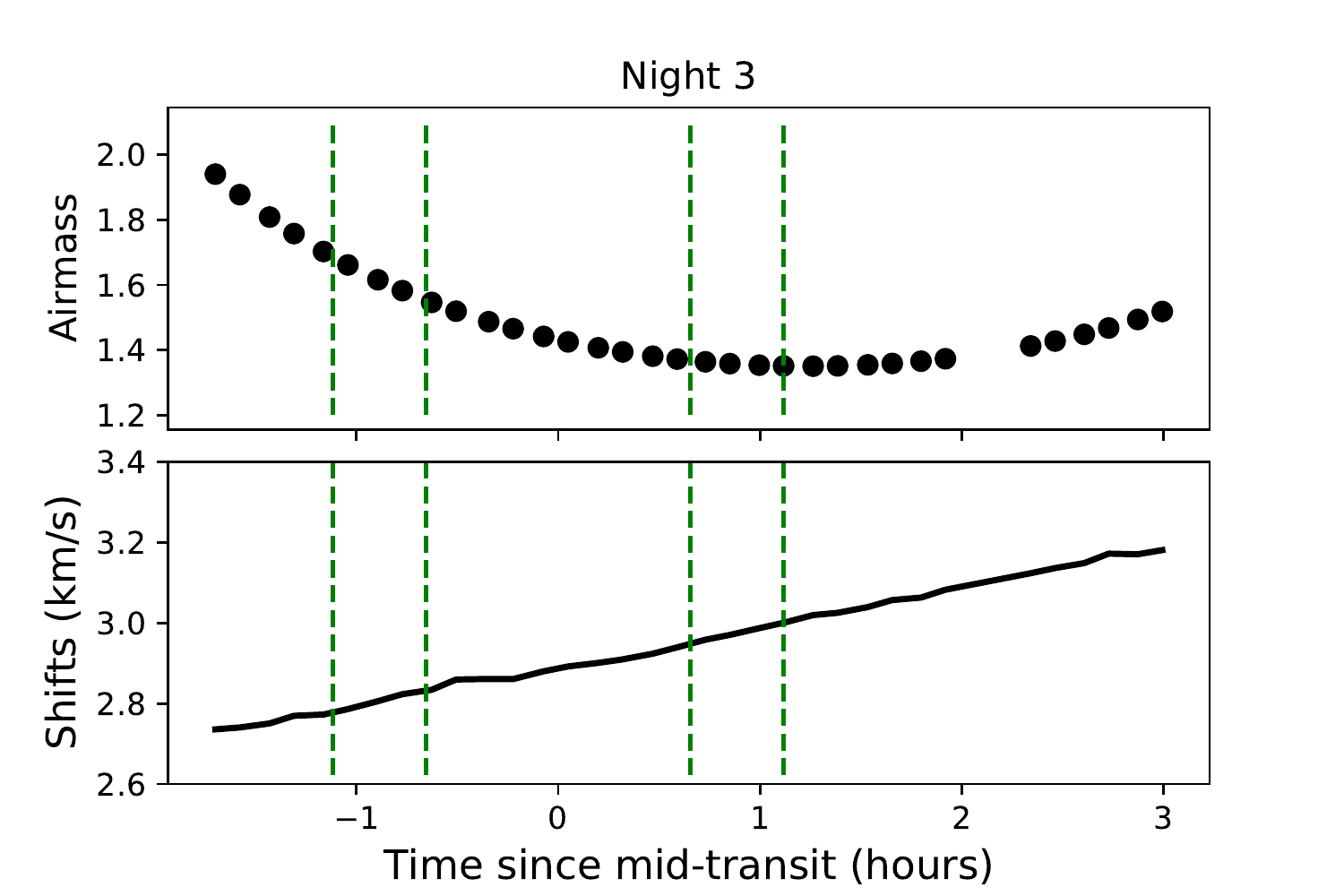}
    \caption{Same as Fig.~\ref{fig:airmass_and_alignment} but for nights 2 ({\em top}) and 3 ({\em bottom}).}
    \label{fig:am_shift_appen}
\end{figure}

\begin{figure}
    \centering
    \includegraphics[width=0.95\columnwidth]{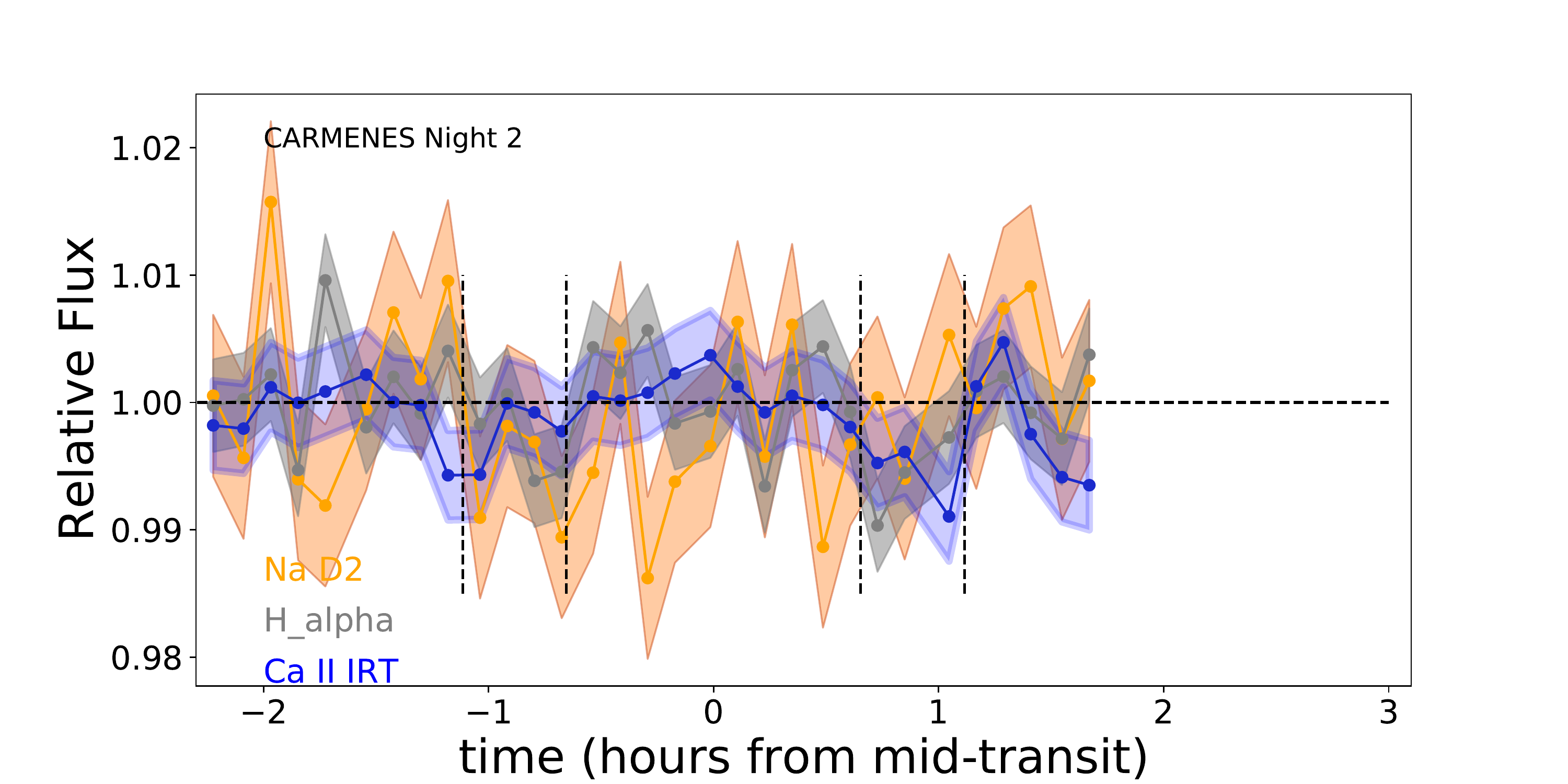}
    \includegraphics[width=0.95\columnwidth]{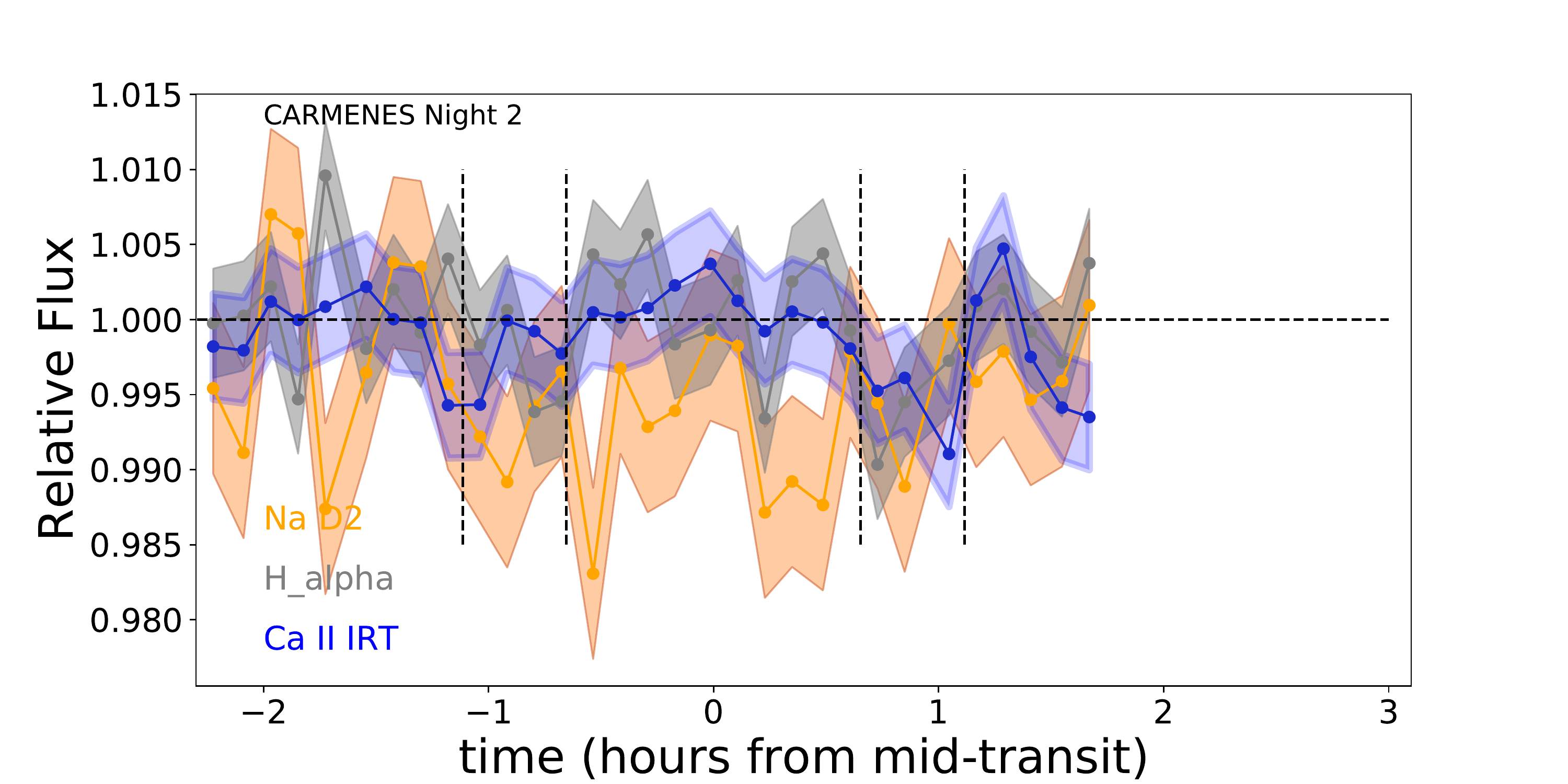}
    \includegraphics[width=0.95\columnwidth]{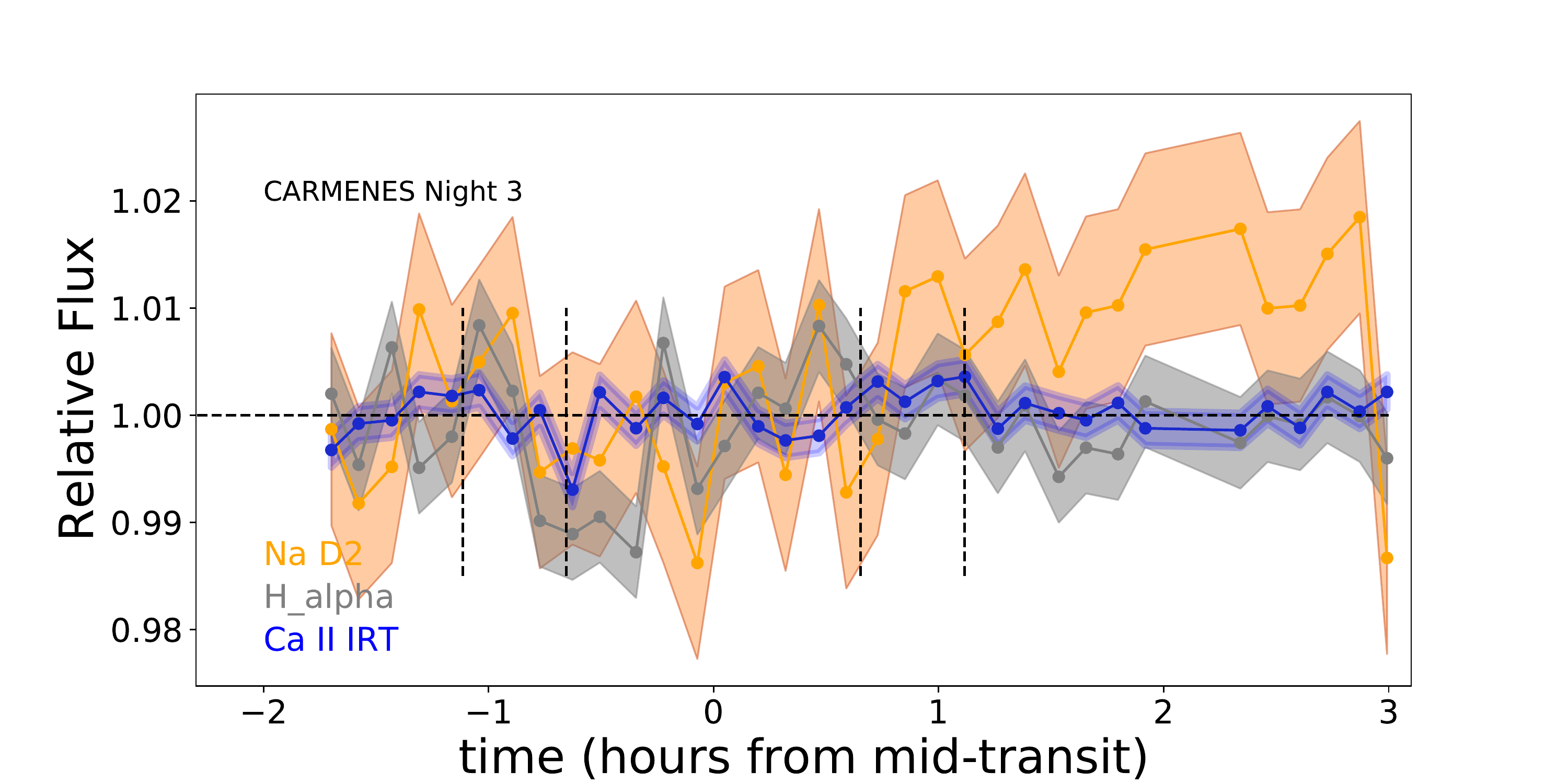}
    \includegraphics[width=0.95\columnwidth]{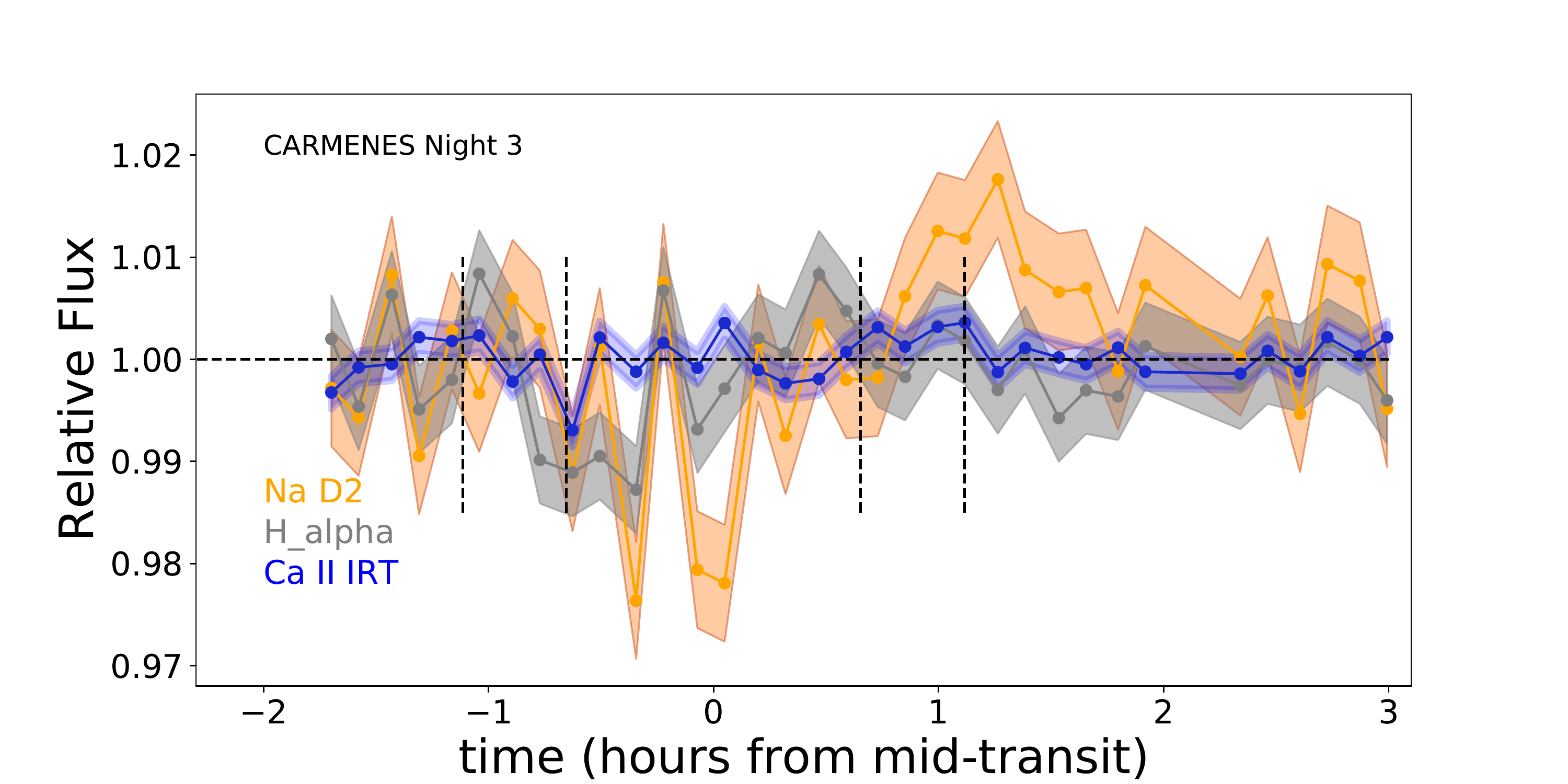}
    \caption{Same as Fig.~\ref{fig:chrom_activity} but for nights 2 ({\em first and second panels}) and 3 ({\em third and fourth panels}).}
    \label{fig:lc_appen}
\end{figure}

\begin{figure*}
    \centering
    \includegraphics[scale = 0.26]{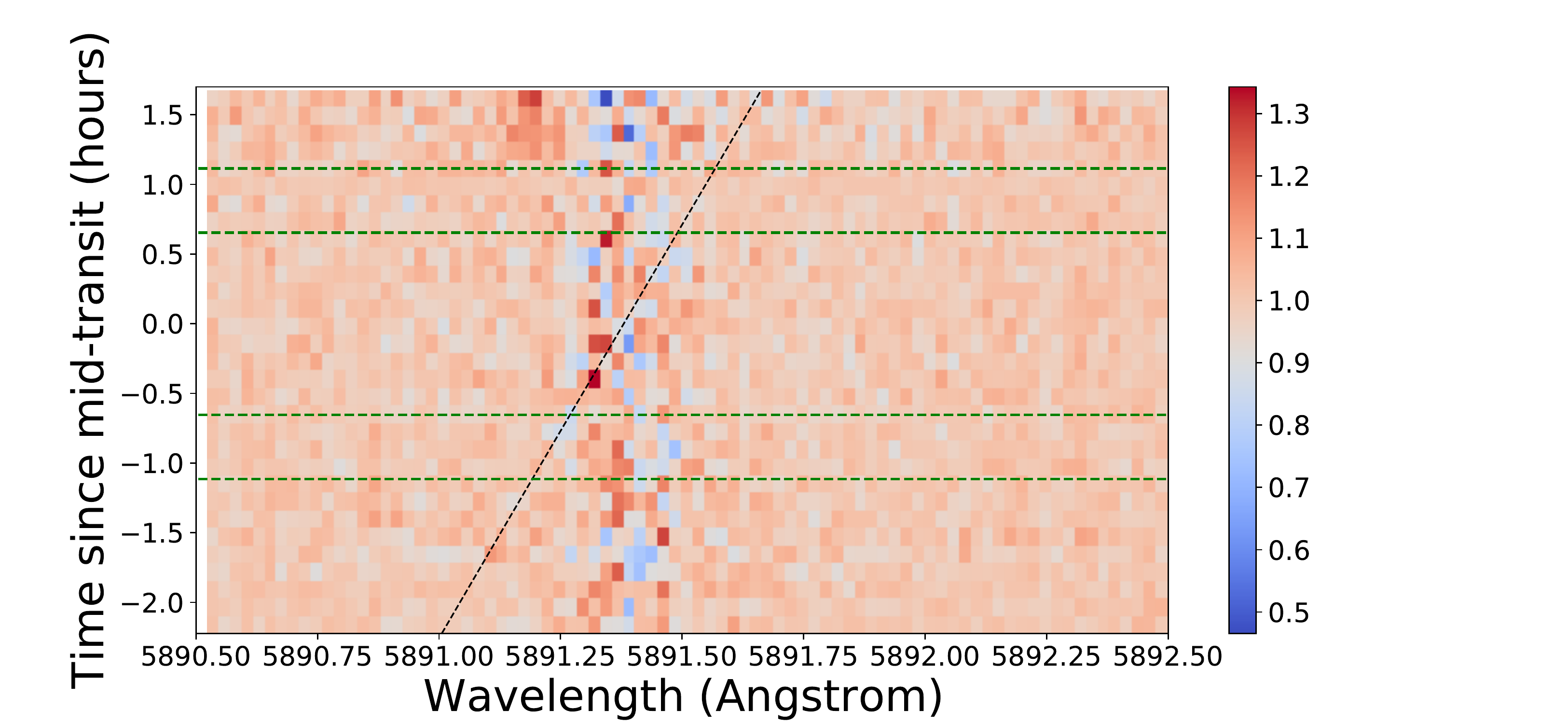}
    \includegraphics[scale = 0.26]{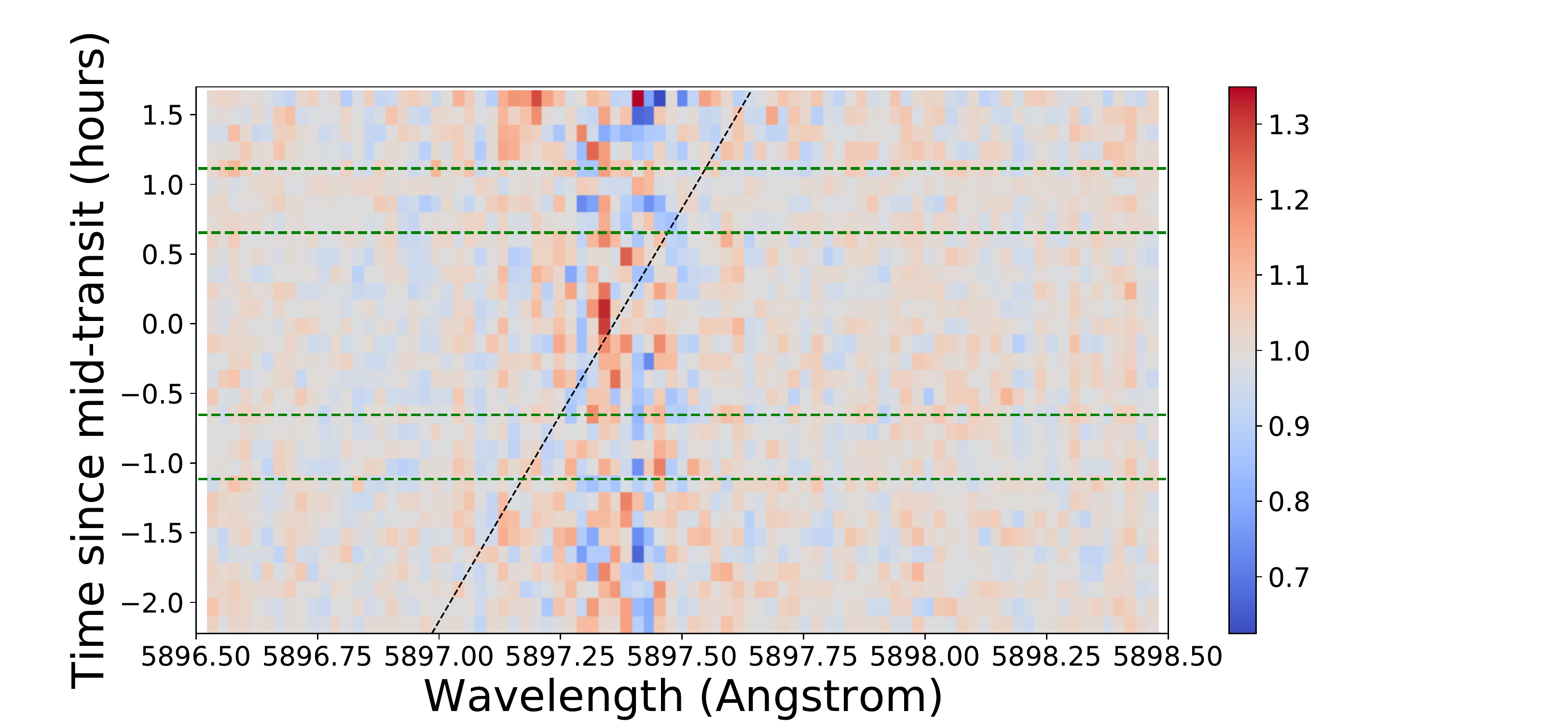}     
    \includegraphics[scale = 0.26]{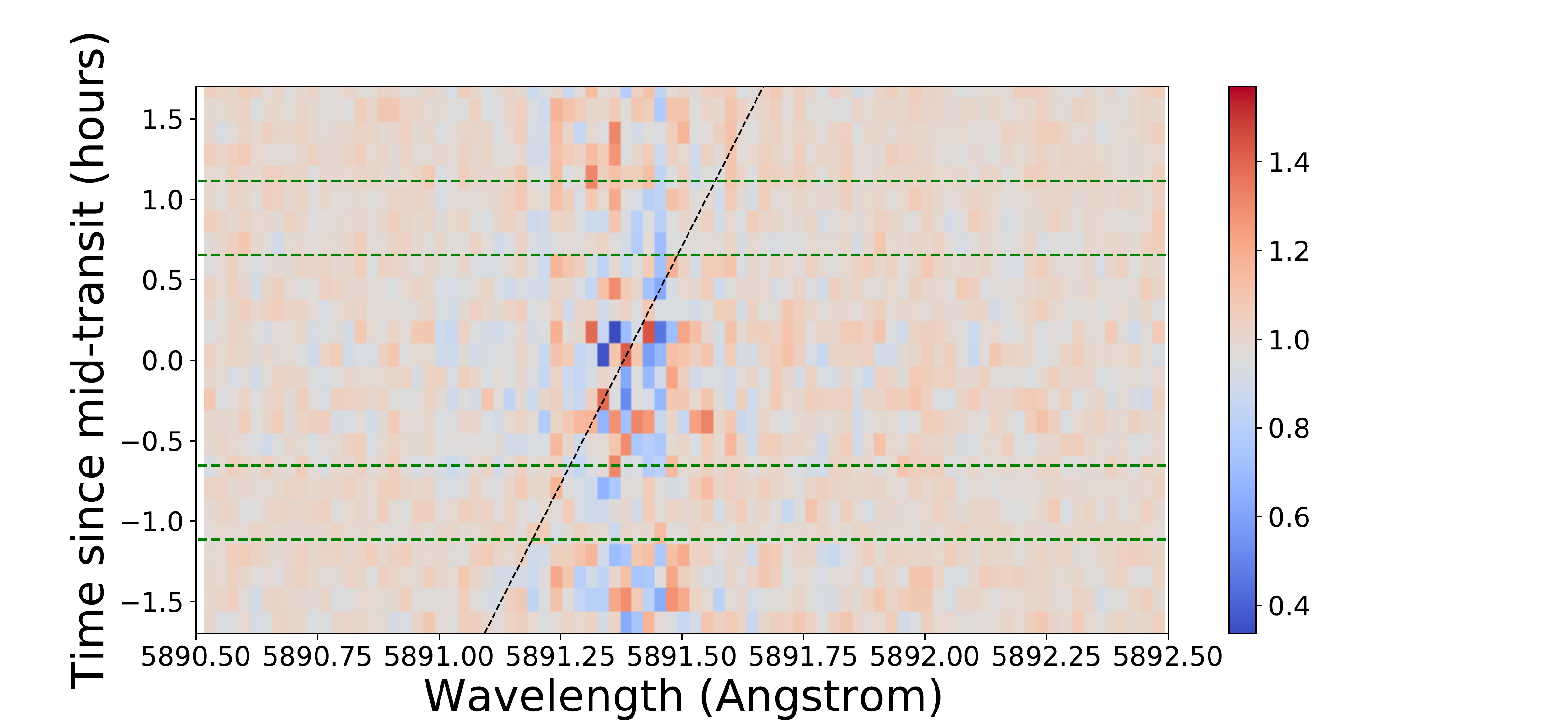}
    \includegraphics[scale = 0.26]{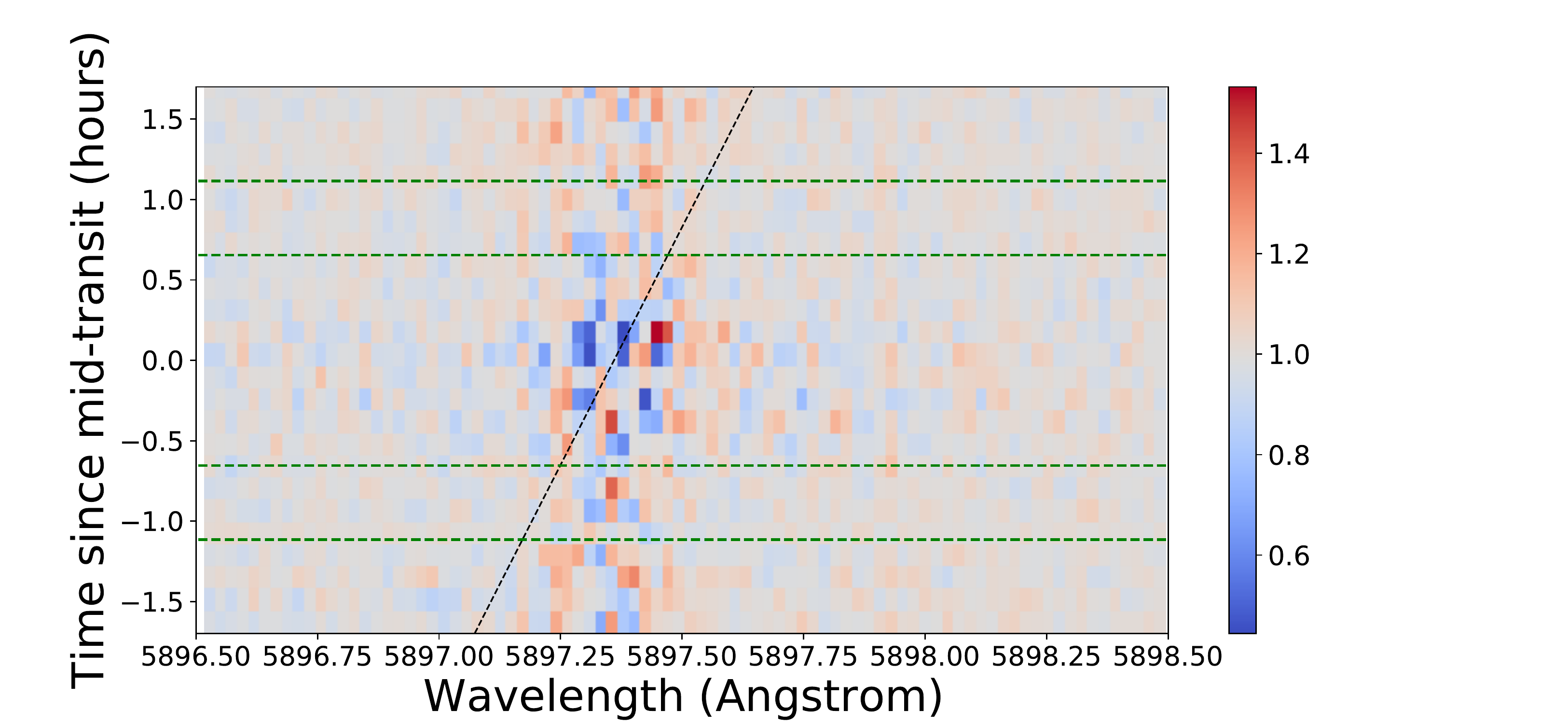} 
    \caption{Two-dimensional map of residuals (same as Fig.~\ref{fig:matrix_plot}) for both the D$_{2}$ ({\em left panels}) and D$_{1}$ ({\em right panels}) lines for nights 2 ({\em top}) and 3 ({\em bottom}).}
    \label{fig:matrix_plot_app}
\end{figure*}

\begin{figure*}
    \centering
    \includegraphics[width=0.85\textwidth]{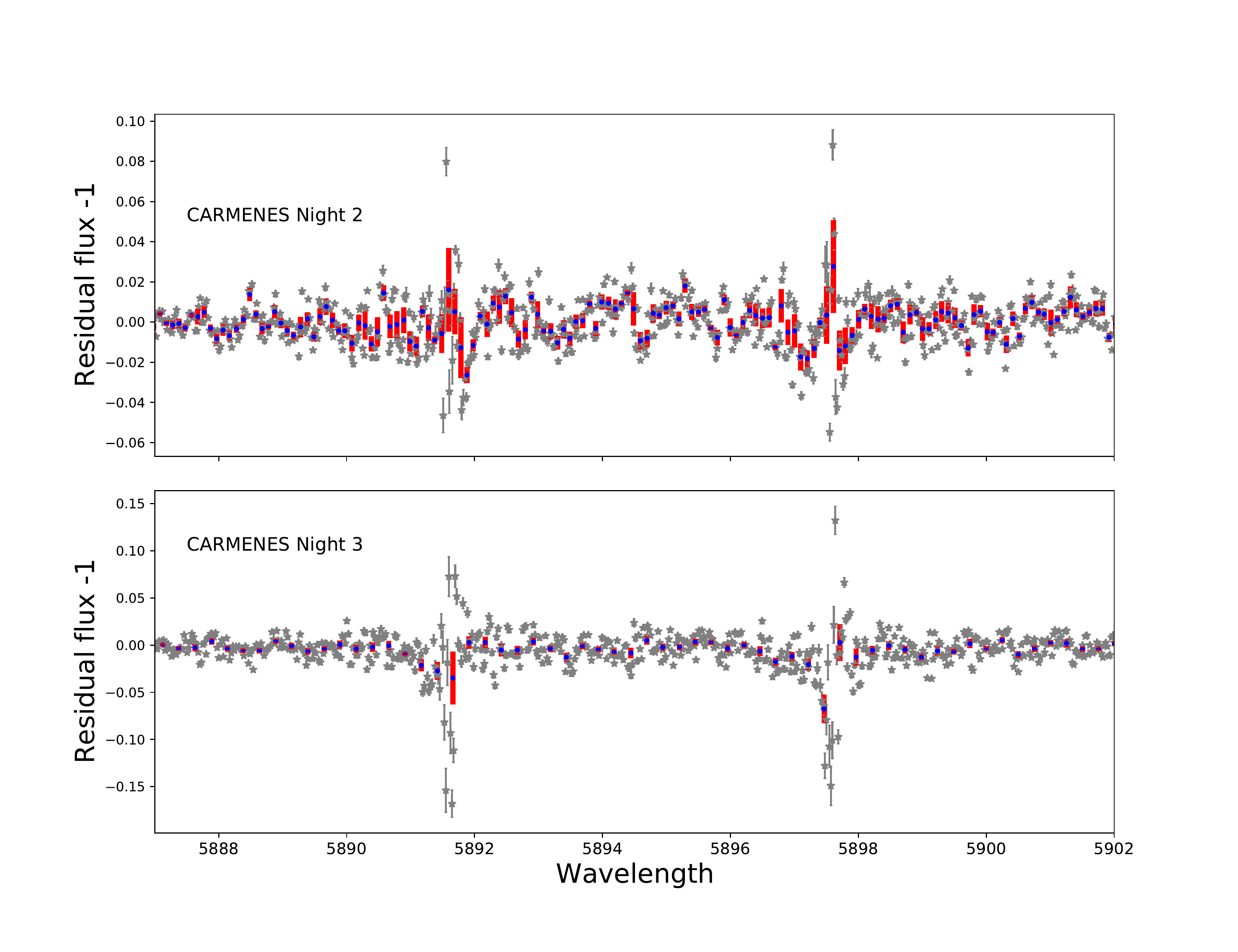}
    \caption{Same as middle panel of Fig.~\ref{fig:TS_Na}, but for nights 2 ({\em top}) and 3 ({\em bottom}).}
    \label{fig:TS_Na_app}
\end{figure*}

In our atmospheric analysis, we did not consider the data from the second and third nights of observations with CARMENES. 
As mentioned in Sect.~\ref{sec:obs_carmenes}, these spectra are highly contaminated by the telluric emission features in the sodium region on the second night, and the data quality and S/N are not sufficient in the visible channel for both nights.
However, for reference and to present stellar chromospheric activity in the cores of H${\alpha}$ and the Ca~{\sc ii} IRT, we show the results of these nights in this section. 
Figure~\ref{fig:am_shift_appen} shows the airmass and the spectral shifts, Fig.~\ref{fig:lc_appen} the time evolution of the cores of the activity indicators H${\alpha}$ and Ca~{\sc ii} IRT, Fig.~\ref{fig:matrix_plot_app} the residual maps around both sodium D-lines, and finally Fig.~\ref{fig:TS_Na_app} the transmission spectrum around the sodium features on each night. The residual map (Fig.~\ref{fig:matrix_plot_app}, bottom panels) and the transmission spectrum (Fig.~\ref{fig:TS_Na_app}, bottom panel) on the third night show some tentative signatures of Na absorption, albeit at low significance, due to the relatively large scatter of the data points around the sodium lines.

\section{Gaussian toy model}
\label{sec:appen_Gauss}

\begin{table*}
\centering
\caption{\label{tab:toy_Gaussian} Parameters of the Gaussian toy model (night 1).}
\begin{tabular}{l l l cc}
    \hline
    \hline
\noalign{\smallskip}
Parameter & Symbol (unit) & Prior & \multicolumn{2}{c}{Posterior} \\
~ & ~ &  ~ & Na~{\sc i} D$_{1}$ & Na~{\sc i} D$_{2}$ \\
\noalign{\smallskip}
    \hline
\noalign{\smallskip}
Mean (center of Gaussian)  & $\mu$ (\AA) & $\mathcal{U}(5891.21,5891.51)$ & \multicolumn{2}{c}{5891.38 $\pm$ 0.01} \\
width & $\sigma$ (\AA) & $\mathcal{U}(0.0,0.5)$ & 0.14 $\pm$ 0.02 & 0.13 $\pm$ 0.01 \\
Amplitude (in relative flux) & A & $\mathcal{U}(-0.3,0.3)$ & --0.012 $\pm$ 0.003 & --0.032 $\pm$ 0.003\\
Offset$^{a}$ (in relative flux)  & $\Delta F$ & $\mathcal{U}(-0.1,0.1)$ & \multicolumn{2}{c}{0.001 $\pm$ 0.000} \\
Wavelength difference between D$_{1}$ and D$_{2}$ & $\Delta \mu$ (\AA) & { fixed} & \multicolumn{2}{c}{5.974} \\
\noalign{\smallskip}
\hline
\end{tabular}
\tablefoot{
    \tablefoottext{a}{The offset is the difference between the continuum level and null relative flux.}
}
\end{table*}

\begin{figure*}
    \centering
    \includegraphics[width=0.80\textwidth]{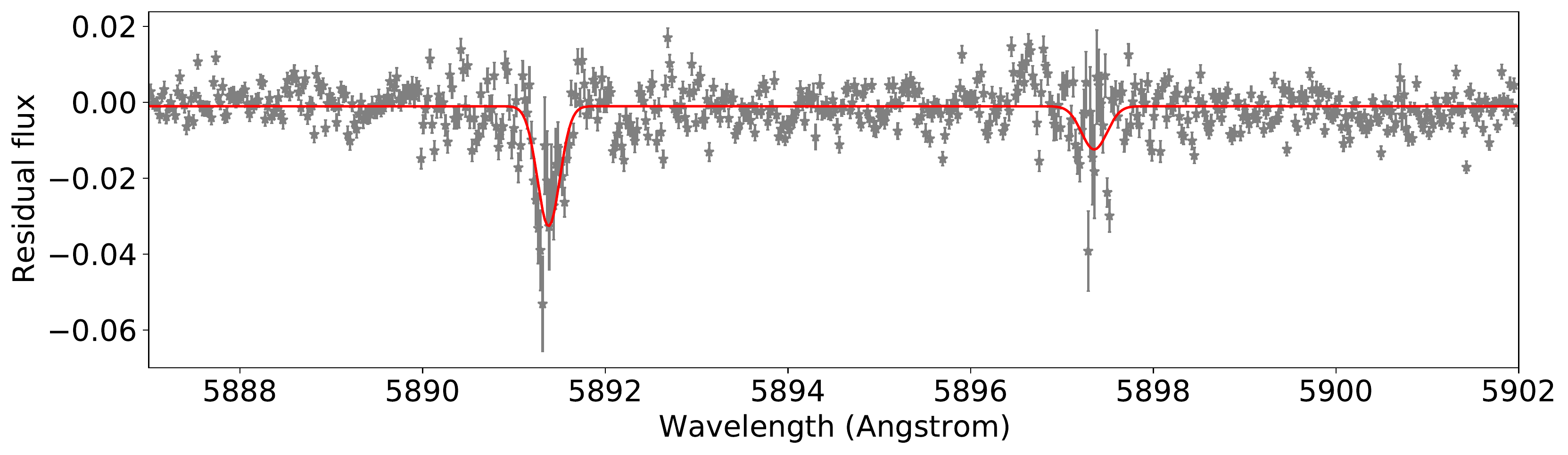}  
    \includegraphics[width=0.80\textwidth]{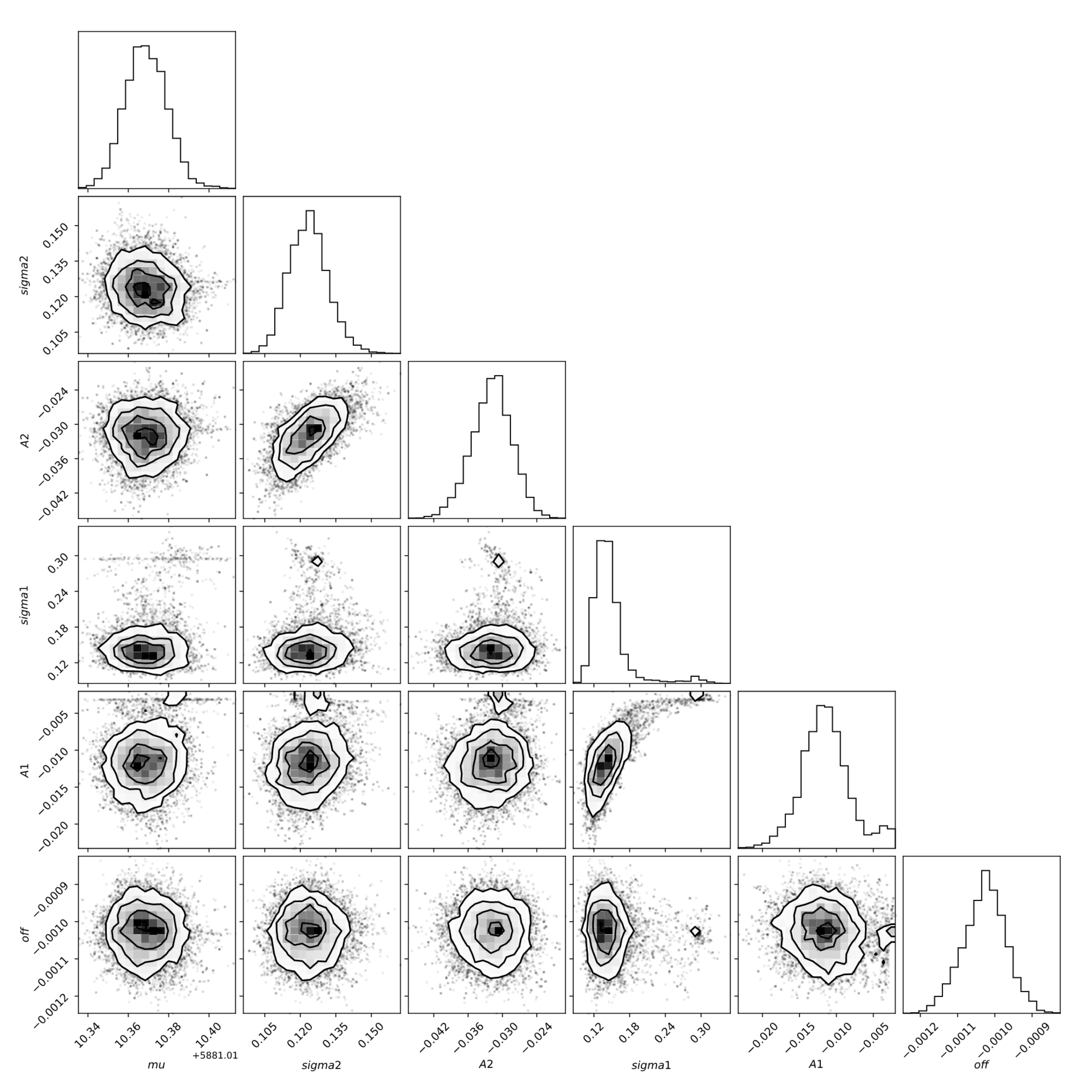}
    \caption{{\em Top panel}: Best-fit combined Gaussian model. 
    {\em Bottom panel}: posterior distribution of the combined Gaussian model parameters.}
    \label{fig:corner_gauss}
\end{figure*}

To measure the absorption signal at the sodium lines and compare it with previous work, we applied a Gaussian toy model to the transmission spectra around the sodium features. This model is a combination of two Gaussian functions, each centered on one of the sodium lines. 
As there is a constant wavelength difference between the cores of each Na D line, we only needed one parameter for the positions of the centers of the Gaussian peaks ($\mu$). 
Here, we selected $\mu$ as the center of D$_{2}$ line.
We employed 200 walkers, with 1000 chains each, where the initial positions were drawn from a Gaussian distribution around our best estimates. 
We imposed uniform priors to all free parameters. 
We allowed a burn-in phase of $\sim$50\,\% of the total chain length, beyond which the MCMC converged. 
The posterior probability distribution was then calculated from the latter 50\,\% of the chain.
The priors and values of the best-fit parameters and their uncertainties are shown in Table~\ref{tab:toy_Gaussian}.
The posterior distribution parameters are shown in Fig.~\ref{fig:corner_gauss}. 

The best model is plotted over the data in the top panel of Fig.~\ref{fig:corner_gauss}, and the correlation diagram of the posterior probability distribution is shown in the bottom panel of the same figure. As Fig.~\ref{fig:corner_gauss} shows, there is a correlation between the width and amplitude parameters of each line. In addition, there is a bi-modality for D$_{1}$, related to the high noise level at this line.

\end{document}